\documentclass{article}
\usepackage[letterpaper,margin=1in]{geometry}
\usepackage{amsfonts}
\usepackage{amsmath}
\usepackage{amssymb}
\usepackage{srcltx}
\usepackage{graphicx}

\def\jump{\vskip 0.05in}
\def\qed{\vrule height 7pt width 3pt depth 0pt}
\newenvironment{proof}{\noindent{\it Proof.} }{\qed\jump}
\newenvironment{sketch}{\noindent{\it Proof sketch.} }{\qed\jump}
\newtheorem{theorem}{Theorem}
\newtheorem{claim}{Claim}

\def\N{{\mathbb{N}}}
\def\R{{\mathbb{R}}}
\def\wh{\widehat}

\newcommand{\newmcommand}[2]{\newcommand{#1}{{\ifmmode {#2}\else\mbox{${#2}$}\fi}}}
\newcommand{\renewmcommand}[2]{\renewcommand{#1}{{\ifmmode {#2}\else\mbox{${#2}$}\fi}}}
\newcommand{\newmcommandi}[2]{\newcommand{#1}[1]{{\ifmmode {#2}\else\mbox{${#2}$}\fi}}}
\newcommand{\newmcommandii}[2]{\newcommand{#1}[2]{{\ifmmode {#2}\else\mbox{${#2}$}\fi}}}
\newcommand{\newmcommandiii}[2]{\newcommand{#1}[3]{{\ifmmode {#2}\else\mbox{${#2}$}\fi}}}

\newfont{\msym}{msbm10}

\newcommand{\paren}[1]{\left({#1}\right)}
\newcommand{\brackets}[1]{\left[{#1}\right]}
\newcommand{\braces}[1]{\left\{{#1}\right\}}

\newcommand{\hf}{H}  

\newcommand{\hyp}{{\cal H}} 

\newmcommand{\M}{\bf M}
\newmcommand{\dM}{\M'}
\newmcommand{\D}{D}
\renewmcommand{\P}{\bf P}
\newmcommand{\Q}{\bf Q}
\newmcommand{\Pt}{\P_t}
\newmcommand{\Qt}{\Q_t}
\newmcommand{\Pstar}{\P^*}
\newmcommand{\Pref}{\tilde{\P}}	
\newmcommand{\Qstar}{\Q^*}
\newmcommand{\Pa}{\overline{\P}}
\newmcommand{\Qa}{\overline{\Q}}
\newmcommandi{\trans}{{#1}^{\rm T}}
\newmcommand{\mhx}{\M(h,x)}
\newmcommand{\mxh}{\dM(x,h)}
\newmcommand{\mpq}{\M(\P,\Q)}
\newmcommand{\mpsq}{\M(\Pstar,\Q)}
\newmcommand{\mpsqt}{\M(\Pstar,\Qt)}
\newmcommand{\mptqt}{\M(\Pt,\Qt)}
\newmcommand{\mptq}{\M(\Pt,\Q)}
\newmcommand{\mpqt}{\M(\P,\Qt)}

\newmcommand{\sumt}{\sum_{t=1}^T}
\newmcommand{\sumin}{\sum_{i=1}^n}
\newmcommand{\delt}{\Delta_T}

\newmcommand{\hyps}{\hyp}
\newmcommand{\predt}{\hat{y}_t}
\newmcommandii{\prob}{\Pr_{#1}\brackets{{#2}}}

\newmcommand{\hfin}{\hf}

\begin{document}
\title{A new Hedging algorithm and its application to inferring latent random variables}

\author{Yoav Freund and Daniel Hsu \\
{\tt \{yfreund,djhsu\}@cs.ucsd.edu}}

\maketitle

\newcommand{\vp}{\vec{p}}
\newcommand{\dt}{\Delta t}

\newcommand{\ctp}[2]{p_{#1}\paren{#2}}   
\newcommand{\ctg}[2]{g_{#1}\paren{#2}}  
\newcommand{\ctga}[1]{g_A\paren{#1}}     
\newcommand{\ctR}[2]{R_{#1}\paren{#2}}  
\newcommand{\ctr}[2]{r_{#1}\paren{#2}}
\newcommand{\ctd}[2]{d_{#1}\paren{#2}}

\begin{abstract}

  We present a new online learning algorithm for cumulative discounted
  gain. This learning algorithm does not use exponential weights on
  the experts. Instead, it uses a weighting scheme that depends on the
  regret of the master algorithm relative to the experts. In
  particular, experts whose discounted cumulative gain is smaller
  (worse) than that of the master algorithm receive zero weight.
  We also sketch how a regret-based algorithm can be used as an
  alternative to Bayesian averaging in the context of inferring latent
  random variables.

\end{abstract}

\section{Introduction} \label{sec:introduction}

We study a variation on the online allocation problem presented by
Freund and Schapire in~\cite{FreundSc97}. Our problem varies from the
original in that we use {\em discounted} cumulative loss instead of
regular cumulative loss. Specifically, we consider the following
iterative game between a {\em hedger} and {\em Nature}.

In this setting, there are $N$ actions (e.g.~strategies, experts) indexed
by $i$. The game between the hedger and Nature proceeds in iterations
$j=0,1,2,\ldots$. In the $j$th iteration:
\begin{enumerate}
\item The hedger chooses a distribution $\{p_i^j\}_{i=1}^N$ over the
  actions, where $p_i^j \geq 0$ and $\sum_{i=1}^N p_i^j = 1$.
\item Nature associates a gain $g_i^j \in [-1,1]$ with action $i$.
\item The gain of the hedger is $g^j_A = \sum_{i=1}^N p_i^j g_i^j$.
\end{enumerate}
We define the {\em discounted total gain} as follows. The initial
total gain is zero $G_i^0 = 0$. The total gain for action $i$ at the start
of iteration $j+1$ is defined inductively as:
\[
G_i^{j+1} \doteq (1 - \alpha) G_i^j + g_i^j
\]
for some fixed {\em discount factor} $\alpha>0$.
The discounted total loss of the hedger is similarly
defined:
\[
G_A^0=0, \quad G_A^{j+1} \doteq (1 - \alpha) G_A^j + g_A^j~.
\]
We define the {\em regret} of the hedger with respect to action $i$ at the
start of iteration $j$ as
\[
R_i^j \doteq G_i^j - G_A^j
\]
It is easy to see that the regret obeys the following recursion:
\[
R_i^0=0, \quad R_i^{j+1} = (1 - \alpha) R_i^j + g_i^j - g_A^j~.
\]
Our goal is to find a hedging algorithm for which we can show a small uniform
upper bound on the regret, i.e. a small positive real number
$B(\alpha)$ such that $R_i^j \leq B(\alpha)$ for all choices of Nature, all
$i$ and all $j$.

Our new hedging algorithm, which we call {\bf NormalHedge}, uses the
following weighting:
\begin{equation} \label{eqn:Hedge-distribution}
w_i^j \doteq
\begin{cases}
R_i^j \exp \paren{\frac{\alpha \brackets{R_i^j}^2}{8}} & \text{if $R_i^j >
0$} \\
                   0               & \text{if $R_i^j \leq 0$.}
\end{cases}
\end{equation}
The hedging distribution is equal to the normalized weights 
$p_i^j = w_i^j / \sum_{k=1}^N w_k^j$ unless all of the weights
are zero, in which case we use the uniform distribution $p_i^j = 1/N$.

Our main result is that if $\alpha$ is sufficiently small, the
following inequality holds uniformly over all game histories:
\begin{equation*}
{1 \over N} \sum_{i=1}^N \Phi\paren{\sqrt{\alpha} R_i^j} < 2.32
\end{equation*}
where
\[
\Phi\paren{x} =
\begin{cases}
\exp \paren{\frac{x^2}{8}}  & \text{if $x > 0$} \\
                   1               & \text{if $x \leq 0$.}
                   \end{cases}
\]
This implies, in particular, that for any $i$ and $j$,
\[
R_i^j \leq \sqrt{\frac{8 \ln 2.32N}{\alpha}}.
\]

The discount factor $\alpha$ plays a similar role to the number of
iterations in the standard undiscounted cumulative loss framework.
Indeed, it is easy to transform the usual exponential weights algorithms
from the standard framework (e.g.~Hedge \cite{FreundSc97}) to our
present setting (Section~\ref{sec:comparison}). Such algorithms also
enjoy discounted cumulative regret bounds of
\[ R_i^j \leq C \cdot \sqrt{\frac{\ln N}{\alpha}} \]
for some positive constant $C$, but they require knowledge of the number of
actions $N$ to tune a learning parameter. The tuning of NormalHedge does
not have this requirement\footnote{The guarantees afforded to NormalHedge
require $\alpha$ to be sufficiently smaller than $1/\ln N$, but this
restriction is operationally different from needing to know $N$ in
advance.}.

The rest of this paper is organized as follows. In
Section~\ref{sec:drifting-game} we describe the main ideas behind the
construction and analysis of NormalHedge. In
Sections~\ref{sec:comparison} and \ref{sec:hedge} we discuss related work and compare NormalHedge to exponential
weights algorithms. Finally, in Section~\ref{sec:latent} we suggest
how to use NormalHedge to track latent variables and sketch how that
might be used for learning HMMs under the $L_1$ loss.

\section{NormalHedge} \label{sec:drifting-game} 

\subsection{Preliminaries}

NormalHedge and its analysis are based on the potential function
$\Phi(x)$ introduced in Section~\ref{sec:introduction}. Here
we give a slightly more elaborate definition for $\Phi(x)$ that includes a
constant $c$. The potential function is a 
non-decreasing function of $x \in \R$
\begin{equation} \label{eqn:potential}
\Phi(x) \doteq \begin{cases}e^{x^2/2c} & \text{if $x > 0$} \\
                      1        & \text{if $x \leq 0$}
                      \end{cases}
\end{equation}
where $c > 1$. In our current version of NormalHedge, $c=4$. Decreasing
$c$ will improve the bound on the regret; we will also argue that $c$
cannot be decreased to $1$.

The weights assigned by NormalHedge are set proportional to the first
derivative of $\Phi$, i.e.~$w_i^j = \Phi'(R_i^j)$, where
$$ \Phi'(x) = \begin{cases}{x \over c}e^{x^2/2c} & \text{if $x > 0$} \\
                   0                   & \text{if $x \leq 0$.}
                   \end{cases} $$
In our analysis, we will also need to examine the second derivative of
$\Phi$:
$$ \Phi''(x) = \begin{cases}\paren{{1 \over c} + {x^2 \over c^2}}e^{x^2/2c} 
                                       & \text{if $x > 0$} \\
                   0                   & \text{if $x < 0$.}
                   \end{cases} $$
Note that $\Phi''(x)$ has a discontinuity at $x=0$.

\subsection{An intuitive derivation}

The intuition behind the potential function is based on considering
the following strategy for Nature. Suppose there are two types of
actions, {\em good} actions and {\em poor} actions. The gain for each
action on each iteration is chosen independently at random from a
distribution over $\{-1,+1\}$. The distribution for poor actions has
equal probabilities $1/2,1/2$ on the two outcomes, while the
distribution for the good experts is $(1+\gamma)/2$ on $+1$ and
$(1-\gamma)/2$ on $-1$ for some very small $\gamma>0$. Clearly, the
best hedging strategy is to put equal positive weights on the good
actions and zero weight on the poor actions. Unfortunately, the
hedging algorithm does not know at the beginning of the game which
experts are good, so it has to learn these weights online. Assuming
that the number of actions is infinite (or sufficiently large), the
per-iteration gain of the optimal weighting is $\gamma$, which implies
that the discounted cumulative gain of this strategy is $\gamma/\alpha$. 

Consider the regrets of this optimal hedging with respect to the good
actions. It is not hard to show that the expected value of the
discounted cumulative gain of a good action is $\gamma/\alpha$ and
that the variance is approximately $1/\alpha$ (becomes exact as
$\gamma \to 0$). Moreover, if $\alpha \to 0$ this distribution
approaches a {\em normal} distribution with mean $\gamma/\alpha$ and
variance $1/\alpha$. In other words, the distribution of the regrets
of optimal hedging with respect to the good actions is
$(1/Z)\exp(-\alpha R^2/2)$. 

Consider the expected value of the potential function
$\Phi(\sqrt{\alpha}R)$ for this distribution over the regrets. If we
set $c=1$ we find that the product of the probability of the regret
$R$ and the potential for the regret $R$ is a constant independent of
$R$:
$$ \frac1Z \cdot \exp\left(-\frac{\alpha R^2}{2} \right) \cdot \exp\left(\frac{\alpha
R^2}{2} \right) = \frac1Z = \Omega(1). $$
Thus the expected potential is infinite. However, if we set $c$ to be
larger than $1$ then the expected value of the potential function becomes
finite. Thus, roughly speaking, the potential associated with a regret
value is the reciprocal of the probability of that regret value being a
result of random fluctuations. This level of regret is unavoidable. The
design of NormalHedge is based on the goal of not allowing the average
regret to grow beyond this level that is generated by random fluctuations.
Ideally, we would be able to use a potential function with any constant $c$
larger than 1. However, what we are able to prove is that the algorithm
works for $c=4$.

The idea of NormalHedge is to keep the average potential small. It is
therefore natural that the weight assigned to each action is proportional
to the derivative of the potential. Indeed, it is easily checked that the
weights $w_i^j$ defined in Equation~\eqref{eqn:Hedge-distribution} are
proportional to $\Phi'(\sqrt{\alpha}R_i^j)$.
This derivative, however, is best viewed when the hedging game is mapped
into continuous time.

\subsection{The continuous time limit}
Our analysis of NormalHedge is based on mapping the integer time steps
$j=0,1,2,\ldots$ into real-valued time steps $t=0,\alpha,2\alpha,\ldots$
and then taking the limit $\alpha \to 0$. Formally, we redefine
the hedging game using a different notation which uses the real valued
time $t$ instead of the time index $j$. We assume a set of $N$
actions (experts), indexed by $i$. The game between the hedging
algorithm and Nature proceeds in iterations
$t=0,\alpha,2\alpha,\ldots$. At each iteration the
following sequence of actions take place.

\begin{enumerate}
\item The hedging algorithm defines a distribution $\braces{\ctp{i}{t}}_{i=1}^N$ over the
  actions. $\ctp{i}{t} \geq 0;\;\; \sum_{i=1}^N \ctp{i}{t} = 1$.
\item Nature associates a gain $\ctg{i}{t} \in [-\sqrt{\alpha},+\sqrt{\alpha}]$ with action $i$.
\item The gain of the hedger is $\ctga{t} = \sum_{i=1}^N \ctp{i}{t} \ctg{i}{t}$.
\end{enumerate}

We skip the definitions of $G_i(t)$ and $G_A(t)$ as these can become
ill-behaved when $\alpha \to 0$. Instead we define the regret directly:
\[
\ctR{i}{0}=0,\;\; \ctR{i}{t+\alpha} = (1- \alpha)\ctR{i}{t} + \ctg{i}{t} - \ctga{t}~.
\]
Note that this definition of the regret is a scaled version of the
discrete time regret:
\[
\ctR{i}{j\alpha} = \sqrt{\alpha} R_i^j.
\]

We now have the tools needed to prove our main result.
\begin{theorem} \label{thm:main}
There exists a positive constant $C < 2.32$ such that if $\alpha < 1/(800
\ln CN)$, then for any sequence of gains and any iteration $j$
$$ \frac1N \sum_{i=1}^N \Phi\paren{\sqrt{\alpha} R_i^j} < C. $$
\end{theorem}
%
\begin{sketch}
The full proof is given in the appendix, but here we sketch a
continuous-time argument (i.e.~we consider $\alpha \to 0$). The formal,
discrete-time proof shows that it is enough for $\alpha \leq 1/(800 \ln
CN)$.

We want to show that the average potential
\[ \Psi(t) \doteq \frac1N \sum_{i=1}^N \Phi(t) \]
is bounded for all time $t$. Our approach is to show that its
time-derivative
\[ \frac{\partial}{\partial t} \Psi(t) = \lim_{\alpha \to 0} \frac1\alpha
\cdot \frac1N \sum_{i=1}^N \left\{ \Phi(\ctR{i}{t+\alpha}) -
\Phi(\ctR{i}{t}) \right\} \]
becomes non-positive as soon as $\Psi(t)$ is above some constant (recall
that the time steps are in increments of $\alpha$). Since the $\Phi(x)$ is
constant for $x < 0$, we need only consider $i$ such that
$\ctR{i}{t+\alpha} \geq 0$. Ignoring the discontinuity of $\Phi''(x)$ at
$x=0$, Taylor's theorem implies that for some $\rho_i \leq
\max\{\ctR{i}{t}, \ctR{i}{t+\alpha}\}$,
\begin{eqnarray*}
 \sum_{i: \ctR{i}{t} \geq 0}
\Phi(\ctR{i}{t+\alpha}) - \Phi(\ctR{i}{t})
& = & \sum_{i: \ctR{i}{t} \geq 0} \Phi((1-\alpha) \ctR{i}{t} + g_i(t) - g_A(t)) - \Phi(\ctR{i}{t}) \\
& = & \sum_{i: \ctR{i}{t} \geq 0} (-\alpha \ctR{i}{t} + g_i(t) - g_A(t))
\Phi'(\ctR{i}{t}) \\
& & \quad \quad \quad \mbox{} + \frac12 (g_i(t) - g_A(t) - \alpha \ctR{i}{t})^2 \Phi''(\rho_i) \\
& \leq & \sum_{i: \ctR{i}{t} \geq 0} -\alpha \ctR{i}{t} \Phi'(\ctR{i}{t}) + \frac12 (g_i(t) - g_A(t) -
\alpha \ctR{i}{t})^2 \Phi''(\rho_i) \\
& \leq & \sum_{i: \ctR{i}{t} \geq 0} -\alpha \ctR{i}{t} \Phi'(\ctR{i}{t}) +
\frac12 (2\sqrt{\alpha} + \alpha
\ctR{i}{t})^2 \Phi''(\ctR{i}{t} + 2\sqrt{\alpha}).
\end{eqnarray*}
The first inequality uses the fact that the weights are proportional to the
derivatives of the potentials
$$ \sum_{i: \ctR{i}{t} \geq 0} g_i(t) \cdot
\frac{\Phi'(\ctR{i}{t})}{\sum_{j: \ctR{j}{t} \geq 0} \Phi'(\ctR{j}{t})} =
g_A(t), $$
and the second inequality follows because $|g_i(t) - g_A(t)| \leq
2\sqrt{\alpha}$. Now dividing by $\alpha$ and $N$ and taking the limit
$\alpha \to 0$, we have
\begin{eqnarray*}
\frac{\partial}{\partial t} \Psi(t)
& \leq & \lim_{\alpha \to 0} \frac1\alpha \cdot \frac1N \sum_{i: \ctR{i}{t}
\geq 0} \frac12 (2\sqrt{\alpha} + \alpha \ctR{i}{t})^2 \Phi''(\ctR{i}{t} +
2\sqrt{\alpha}) -\alpha \ctR{i}{t} \Phi'(\ctR{i}{t}) \\
& = & \lim_{\alpha \to 0} \frac1N \sum_{i: \ctR{i}{t}
\geq 0} \frac12 (2 + \sqrt{\alpha}\ctR{i}{t})^2 \Phi''(\ctR{i}{t} +
2\sqrt{\alpha}) - \ctR{i}{t} \Phi'(\ctR{i}{t}) \\
& = & \frac1N \left\{ 2 \left( \frac1c + \frac{\ctR{i}{t}^2}{c^2} \right)
\exp(\ctR{i}{t}^2/2c) - \frac{\ctR{i}{t}^2}{c} \exp(\ctR{i}{t}^2/2c) \right\}
\\
& \leq & \frac2c \Psi(t) + \frac1{cN} \sum_{i=1}^N \left( \frac2c - 1 \right)
\ctR{i}{t}^2 \exp(\ctR{i}{t}^2/2c).
\end{eqnarray*}
If $\Psi(t) \geq B$, then this final RHS is maximized when $R_i(t) \equiv
\sqrt{2c\ln B}$ for all $i$, whereupon
$$ \frac{\partial}{\partial t} \Psi(t) \leq \frac{2B}{c} \left( 1 + \left(
\frac2c - 1\right) c \ln B \right). $$
This is non-positive for sufficiently large $B$ and $c \geq 2 + 1/\ln B$.
\end{sketch}

\section{Related work} \label{sec:comparison}

\subsection{Relation to other online learning algorithms}

The Hedge algorithm~\cite{FreundSc97}, as well as most of the work on
online learning algorithms is based on exponential weighting, where
the weight assigned to an expert is exponential in the cumulative loss
of that expert. NormalHedge uses a very different weighting
scheme. The most important difference is that the weight of an expert
depends on the regret of the master algorithm relative to that expert,
rather than just on the loss of the algorithm. In particular, experts
whose discounted cumulative loss is larger than that of the master
algorithm receive zero weight. We expand on the comparison of NormalHedge
to Hedge in Section~\ref{sec:hedge}.

The starting point for the derivation and analysis of NormalHedge is
the Binomial Weights algorithm of Cesa-Bianchi et
al~\cite{CesabianchiFrHeWa96}. The Binomial weights algorithm is an
algorithm for a restricted version of the experts prediction
problem~\cite{LittlestoneWa94,CesabianchiFrHeHaScWa97}. In this version
sequence to be predicted is binary and all of the predictions are also
binary. The Binomial Weights algorithm is analyzed using a type of
{\em chip game}. In this game each expert is represented as a chip, at
each iteration each chip has a location on the integer line. The
position of the chip corresponds to the number of mistakes that were
made by the expert. The a-priori assumption is that there is at least
one experts which makes at most $k$ mistakes, and the goal is to
define a rule for combining the experts predictions in a way that
would minimize the maximal number of mistakes of the master expert.

The chip game analysis leads naturally to the definition of the {\em
  potential function} and the evolution of this potential function
from iteration to iteration yields the Binomial Weights algorithm.  A
closely related notion of potential was used in the Boost-by-Majority
algorithm. The chip-game analysis was extended by Schapire's work on
drifting games~\cite{Schapire01} and by Freund and Opper's work on drifting
games in continuous time~\cite{FreundOp02}. NormalHedge naturally extends
the continuous time drifting games to a setting in which one seeks to
minimize discounted loss.

\subsection{Relation to switching and sleeping experts}

The use of discounted cumulative loss represents an alternative to the
``switching experts'' framework of Warmuth and
Herbster~\cite{HerbsterWa98}. If the best expert changes at a rate of
$O(\alpha)$, then NormalHedge
will switch to the new best expert because the losses that occurred more
than $1/\alpha$ iterations ago make a small contribution to the discounted
total loss.

A useful extension of NormalHedge is to using experts that can
abstain, similar to the setup studied in ~\cite{FreundScSiWa97}. To do this we assume
that each expert $i$, at each iteration $j$, outputs a confidence
level $0 \leq c \leq 1$. Instead of using the vector
$\{p_i^j\}_{j=1}^N$ the hedger uses the vector $\{p_i^j
c_i^j/Z^j\}_{j=1}^N$ where $Z = \sum_{i=1}^N p_i^j c_i^j$. The gain
$g_i^j$ of action $i$ at iteration $j$ is replaced by $c_i^jg_i^j$,
and the discounted cumulative gain and the discounted cumulative
regret change in the corresponding way. The bounds on the average
potential transfer without change. This allows an expert to abstain
from making a prediction. By setting $c_i^j=0$ the expert effectively
removes itself from the pool of experts used by the hedger. It also
avoids suffering any loss. However, an expert cannot always abstain,
because then it's discounted cumulative gain will be driven to zero by
the discount factor.
We will use this extension in Section~\ref{sec:latent}.

\section{Comparison of NormalHedge and Hedge} \label{sec:hedge}

\subsection{Discounted regret bound for Hedge}

To ease the comparison, we first recast the Hedge algorithm
\cite{FreundSc97} into our current framework with discounted gains. The
weights used by Hedge are
$$ w_i^j \doteq \exp(\eta G_i^j) $$
where $G_i^j$ is the discounted cumulative gain of action $i$ at the
start of iteration $j$, and $\eta > 0$ is the learning rate parameter.
When written recursively as
$$ w_i^{j+1} = \exp\left(\eta ((1-\alpha) G_i^j + g_i^j) \right)
\propto \left( w_i^j \right)^{1-\alpha} \exp(\eta g_i^j), $$
we see that the effect of discounting is a dampening of the previous
weights $w_i^j$ prior to the usual multiplicative update rule.

Fix any iteration $j$ and define the adjusted cumulative gain of action $i$
at the start of iteration $k$ to be
\[ \wh G_i^k = \sum_{s=1}^{k-1} (1-\alpha)^{j-1-s} g_i^s \]
with $G_i^0 = 0$. The gain of Hedge in iteration $k$ is
\[ g_A^k =
\frac{\sum_{i=1}^N w_i^k g_i^k}{\sum_{i=1}^N w_i^k} 
= \frac{\sum_{i=1}^N e^{\eta \wh G_i^k}
g_i^k}{\sum_{i=1}^N e^{\eta \wh G_i^k}}
\]
and the adjusted cumulative gain of Hedge at the start of iteration $k$ is
\[ \wh G_A^k = \sum_{s=1}^{k-1} (1-\alpha)^{j-1-s} g_A^s. \]
Then the discounted cumulative regret to action $i$ at the start of
iteration $j$ is $\wh G_i^j - \wh G_A^j$.

We analyze the (log of the) ratios $W_k / W_{k-1}$, where
\[ W_k = \sum_{i=1}^N e^{\eta \wh G_i^k} \]
and $W_0 = N$. We lower bound $\ln(W_j / W_0)$ as
\[ \ln \frac{W_j}{W_0} = \ln \sum_{i=1}^N e^{\eta \wh G_i^j} - \ln N \geq
\ln e^{\eta \wh G_i^j} - \ln N = \eta \wh G_i^j - \ln N \]
(for any $i$), and we upper bound it as
\begin{align*}
\ln \frac{W_j}{W_0}
& = \sum_{k=1}^{j-1} \ln \frac{W_j}{W_{j-1}} \\
& = \sum_{k=1}^{j-1} \ln \frac{\sum_{i=1}^N e^{\eta \wh G_i^{k-1}} e^{\eta
(1-\alpha)^{j-1-k} g_i^k}}{\sum_{i=1}^N e^{\eta \wh G_i^{k-1}}} \\
& \leq \sum_{t=1}^T \eta \cdot \frac{\sum_{i=1}^N e^{\eta \wh G_i^{k-1}}
(1-\alpha)^{j-1-k} g_i^k}{\sum_{i=1}^N e^{\eta G_i^{k-1}}} +
\frac{\eta^2}{8} \cdot 4(1-\alpha^{2(j-1-k)} \quad \text{(Hoeffding's
inequality)} \\
& = \sum_{k=1}^{j-1} \eta (1-\alpha)^{j-1-k} g_A^k + \frac{\eta^2}{2}
(1-\alpha)^{2(j-1-k)} \\
& = \eta \wh G_A^k + \frac{\eta^2}{2} \cdot \frac{1}{1-(1-\alpha)^2} \\
& = \eta \wh G_A^k + \frac{\eta^2}{4(\alpha - \alpha^2/2)}.
\end{align*}
Therefore, the discounted cumulative regret of Hedge to action $i$ at the
start of any iteration $j$ is
$$ R_i^j = \wh G_i^j - \wh G_A^j \leq \frac{\ln N}{\eta}
+ \frac{\eta}{4(\alpha - \alpha^2/2)}. $$
Choosing $\eta = \sqrt{4(\alpha - \alpha^2/2) \ln N}$ gives
$$ R_i^j \leq \sqrt{\frac{\ln N}{\alpha - \alpha^2/2}}. $$

The regret bound is of the same form as that implied by
Theorem~\ref{thm:main}, indeed, with better leading constants. However,
this bound only holds when $\eta$ is tuned with knowledge of the number of
actions $N$. If instead one sets $\eta = \Theta(\sqrt{\alpha})$
independently of $N$, the bound for Hedge is worse by a factor of
$\Theta(\sqrt{\ln N})$. Furthermore, this setting of $\eta$ is for
optimizing a bound that anticipates the worst-case sequence of gains; when
Nature is not optimally adversarial, then a proper setting of $\eta$ may
require other prior knowledge.

\subsection{Simulations}

\subsubsection{The effect of good experts}

To empirically compare Hedge and NormalHedge, we first simulated the two
algorithms in a scenario similar to that described in
Section~\ref{sec:drifting-game}:
\begin{itemize}
\item The number of experts is $N = 1000$, and the discount parameter is
$\alpha = 0.001$.
\item At any given time, there is a set of $N_G = f \cdot N$ good experts
and $N - N_G$ bad experts. (We varied $f \in \{ 0.001, 0.01, 0.1, 0.5 \}$.)
  \begin{itemize}
  \item With probability $0.5 + \gamma/2$, \emph{every} good expert
  receives gain $+1$; with probability $0.5 - \gamma/2$, \emph{every} good
  expert receives gain $-1$. (We varied $\gamma \in \{ 0.2, 0.4, 0.6, 0.8
  \}$.)
  \item Bad experts receive gain $+1$ and $-1$ with equal probability.
  \end{itemize}
\item Initially, the set of good experts is $\{ 0, 1, \ldots, N_G-1 \}$.
\item After every $1/\alpha$ iterations, the set of good experts shifts
from $\{ i_0, i_0 + 1, \ldots, i_0 + N_G-1 \}$ to $\{ i_0 + N_G, i_0 +
N_G+1, \ldots, i_0 + 2N_G-1 \}$ (with addition modulo $N$).
\end{itemize}
Thus, the set of good experts completely changes every $1/\alpha$
iterations. In each iteration, all good experts receive the same gain,
which is $\gamma$ in expectation. In contrast, the gain of each bad expert
is decided independently with a fair coin.

We tuned the learning rate parameter for Hedge to $\eta = \sqrt{(\alpha -
\alpha^2/2) \ln N}$. For NormalHedge, we varied $c \in \{ 1, 2, 4\}$.
Recall that the regret bound we can show for NormalHedge holds for $c = 2$
as $\alpha \to 0$ (the formal proof is stated with $c = 4$).

\begin{figure}
\begin{center}
\begin{tabular}{cc}
\includegraphics[width=0.43\textwidth]{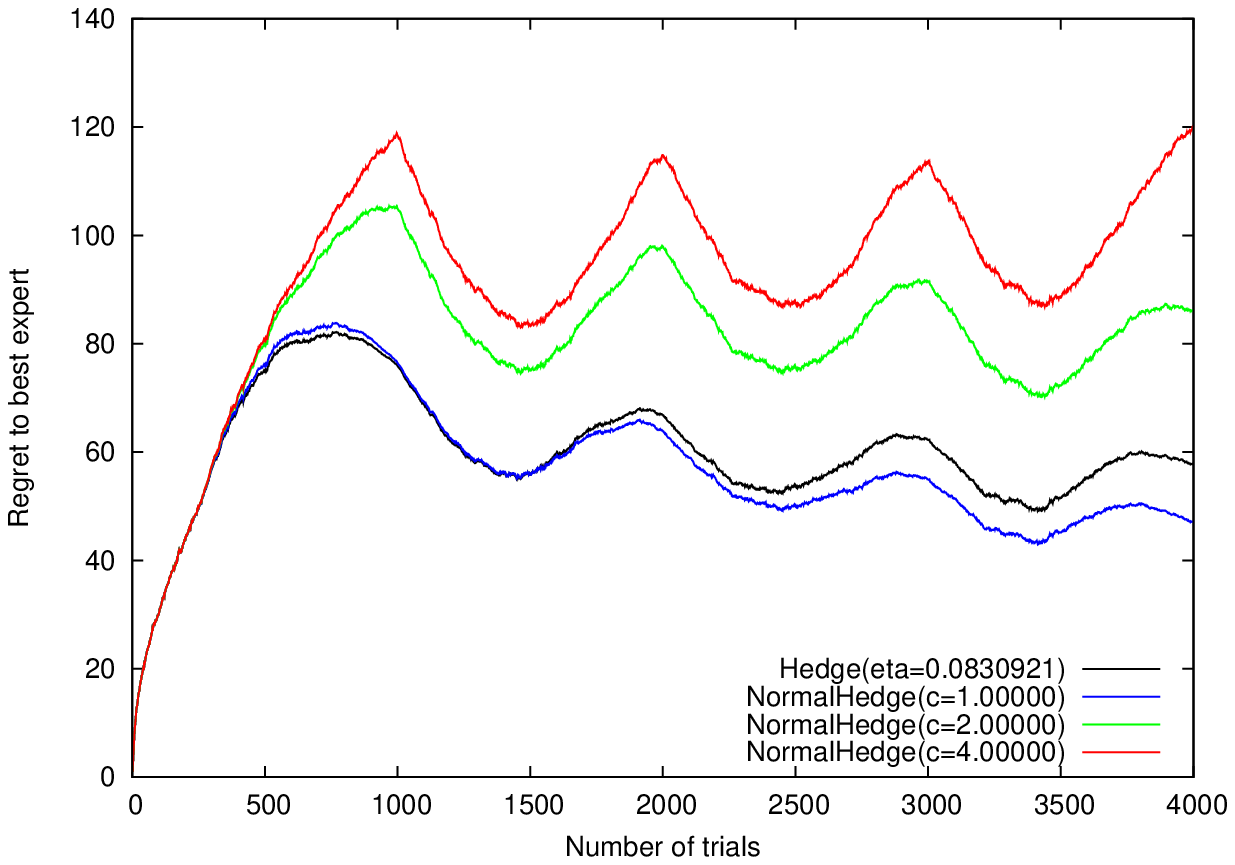} &
\includegraphics[width=0.43\textwidth]{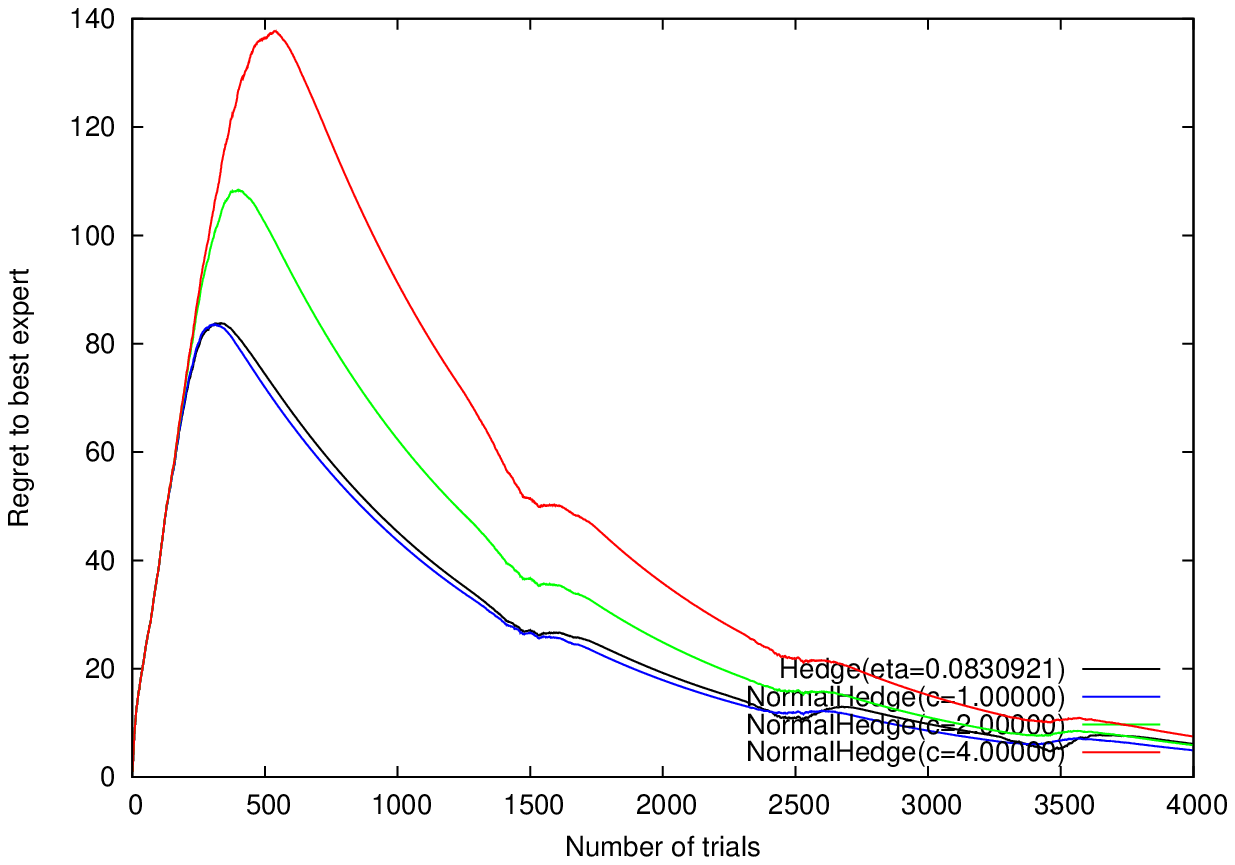} \\
$\gamma = 0.2, f = 0.001$ & $\gamma = 0.4, f = 0.001$ \\
\includegraphics[width=0.43\textwidth]{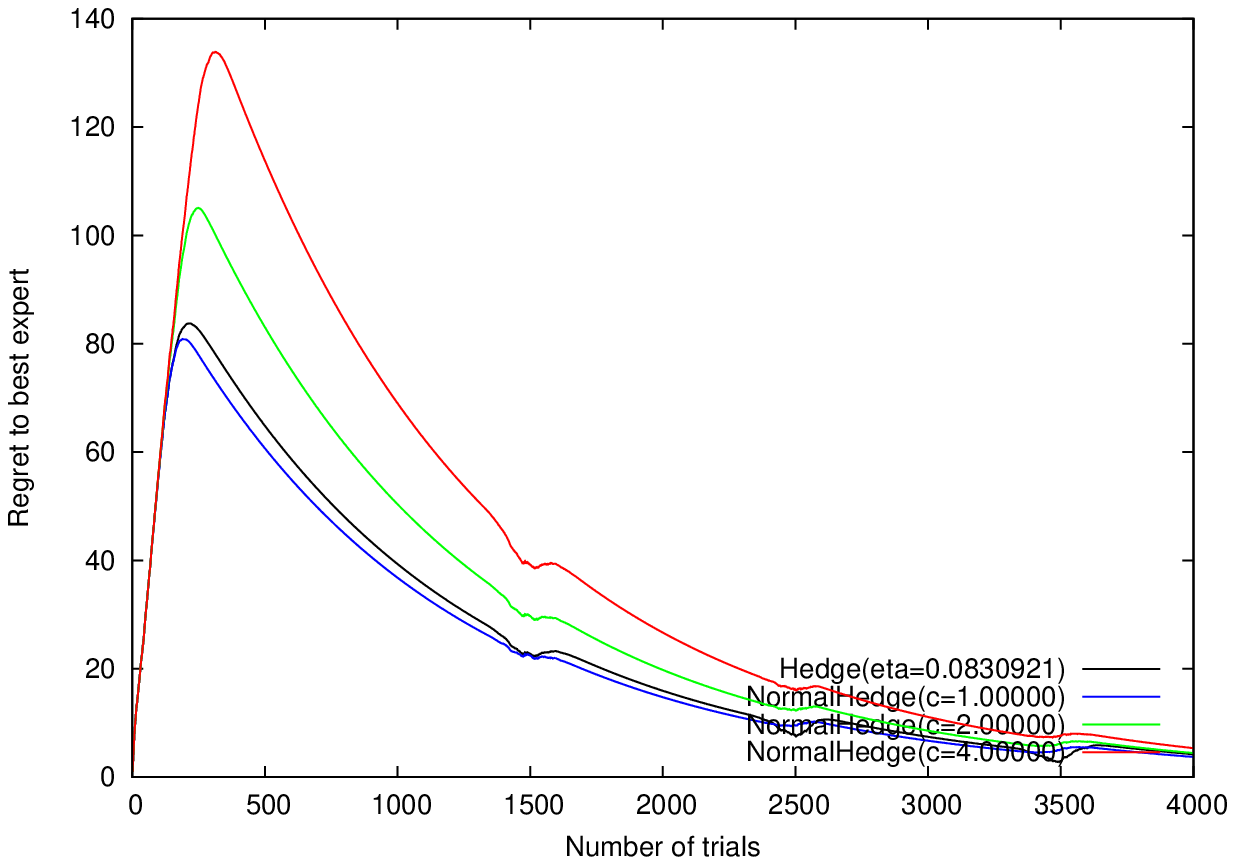} &
\includegraphics[width=0.43\textwidth]{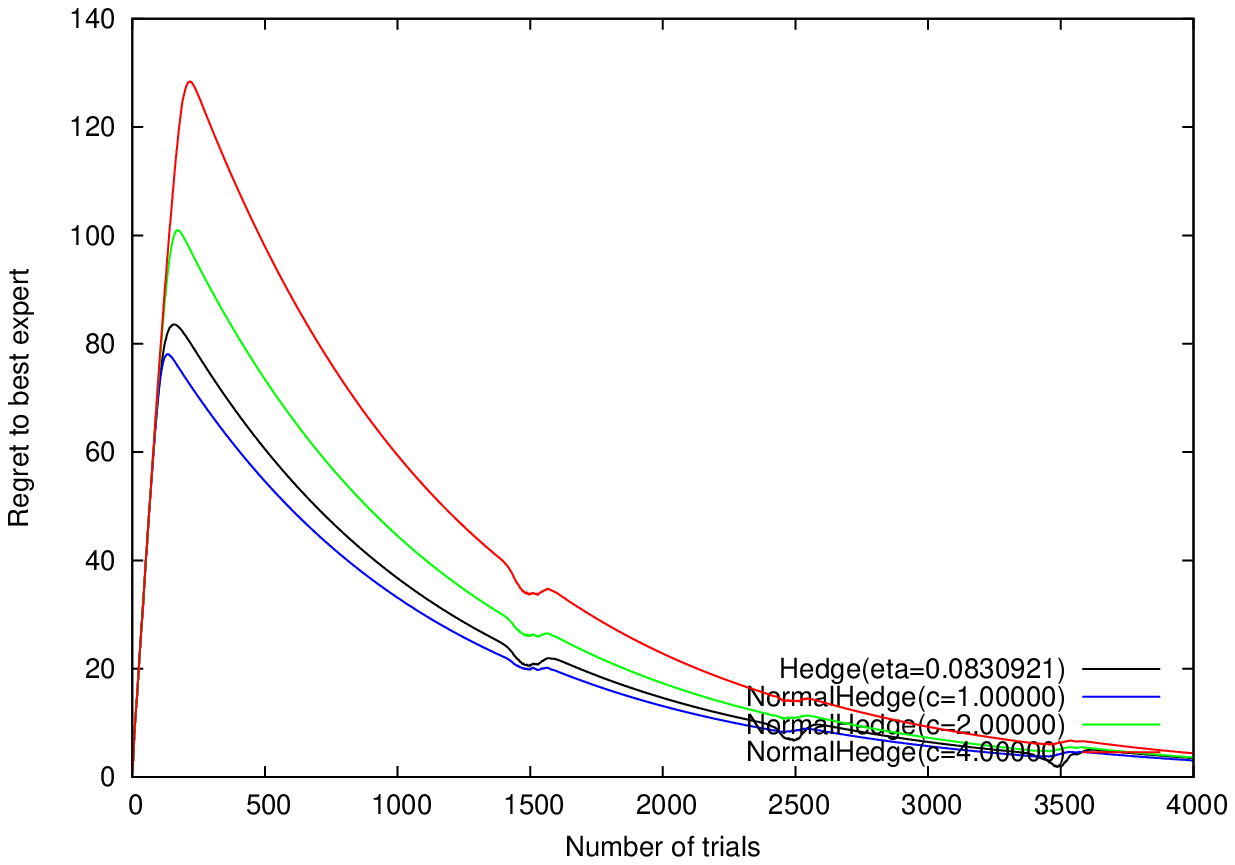} \\
$\gamma = 0.6, f = 0.001$ & $\gamma = 0.8, f = 0.001$ \\
\includegraphics[width=0.43\textwidth]{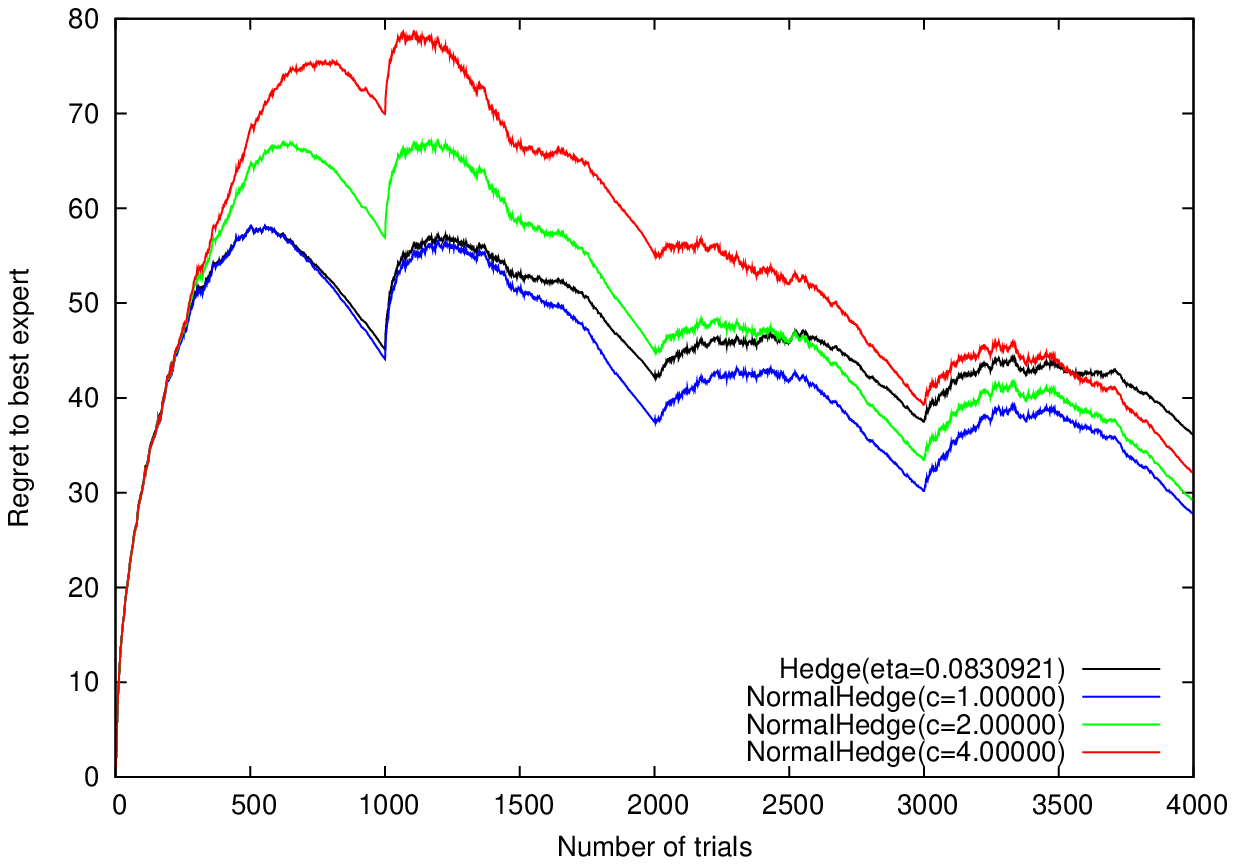} &
\includegraphics[width=0.43\textwidth]{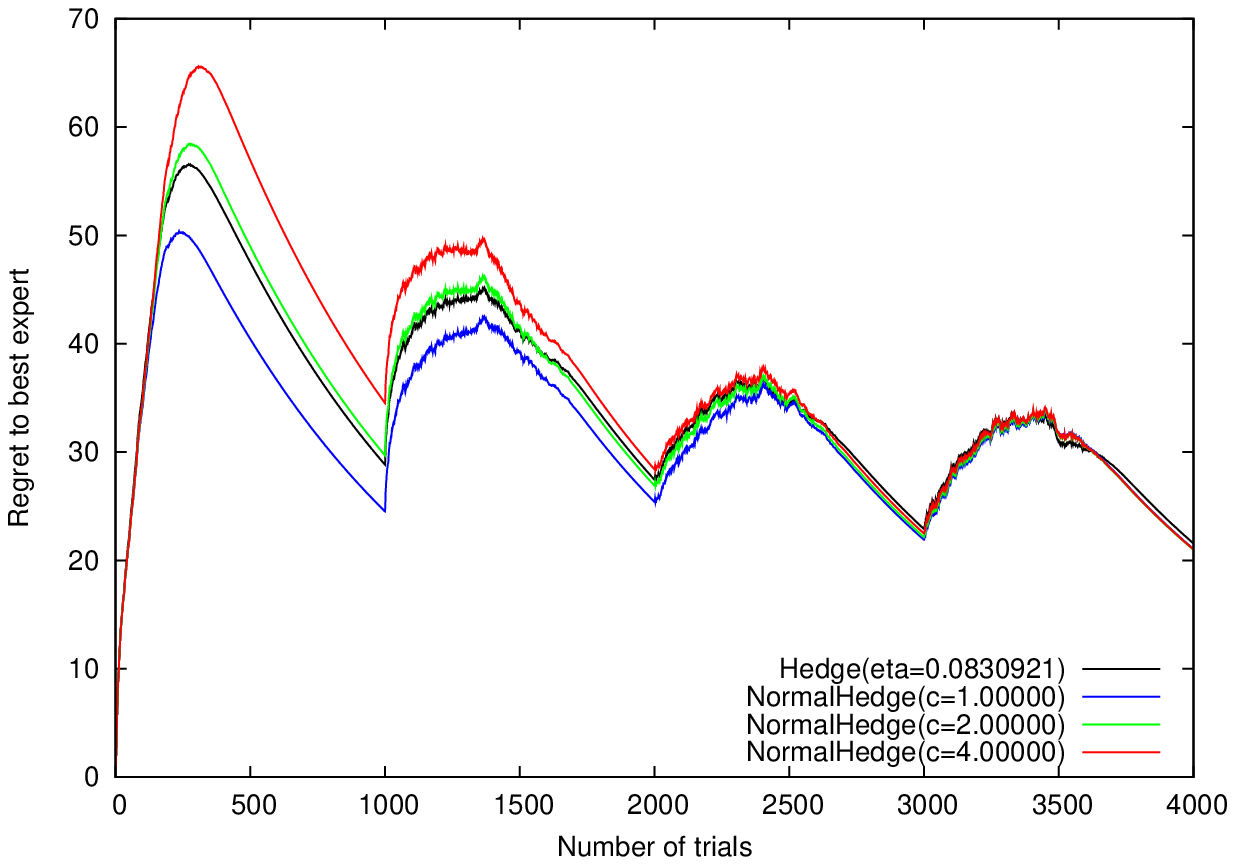} \\
$\gamma = 0.2, f = 0.01$ & $\gamma = 0.4, f = 0.01$ \\
\includegraphics[width=0.43\textwidth]{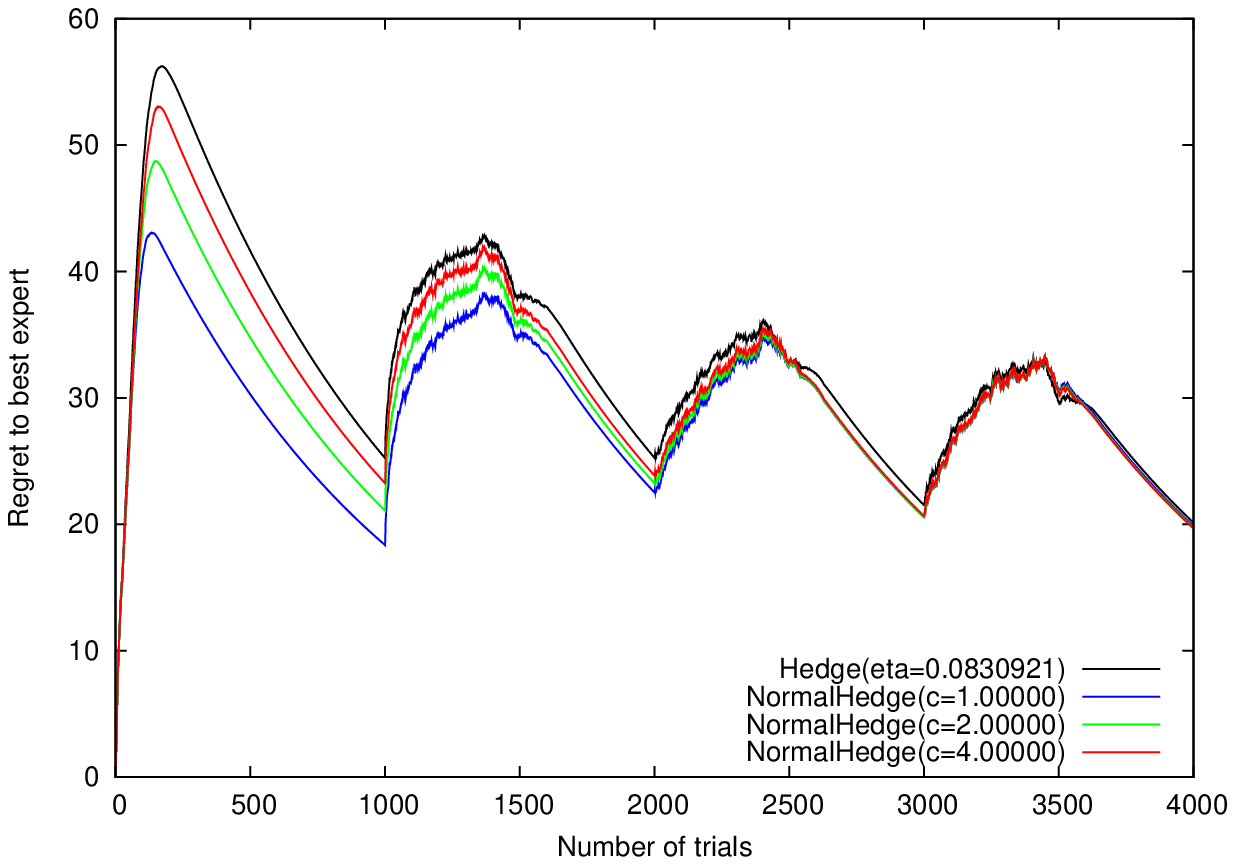} &
\includegraphics[width=0.43\textwidth]{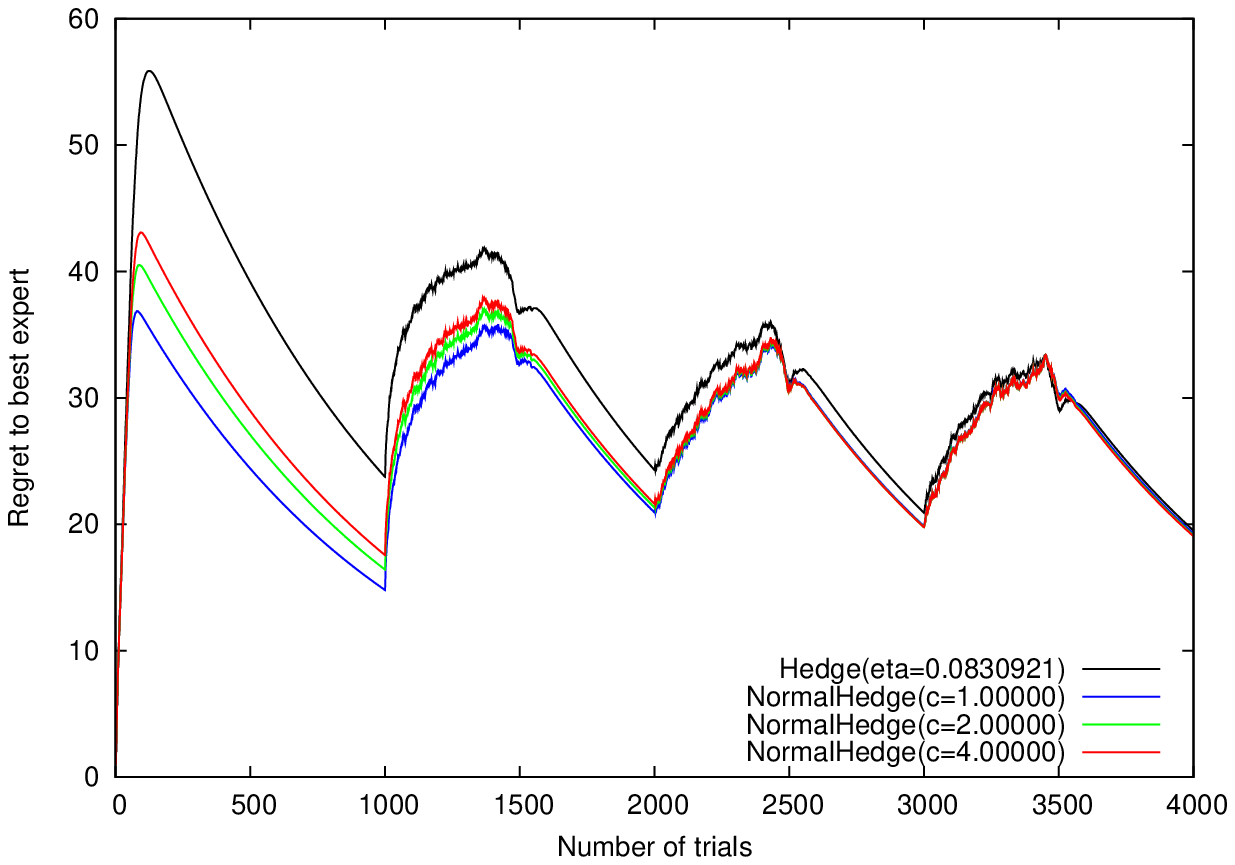} \\
$\gamma = 0.6, f = 0.01$ & $\gamma = 0.8, f = 0.01$
\end{tabular}
\end{center}
\caption{Regrets to the best expert in the first simulation;
$\gamma \in \{ 0.2, 0.4, 0.6, 0.8 \}$ and $f \in \{ 0.001, 0.01 \}$.}
\label{fig:sim1-1}
\end{figure}

\begin{figure}
\begin{center}
\begin{tabular}{cc}
\includegraphics[width=0.43\textwidth]{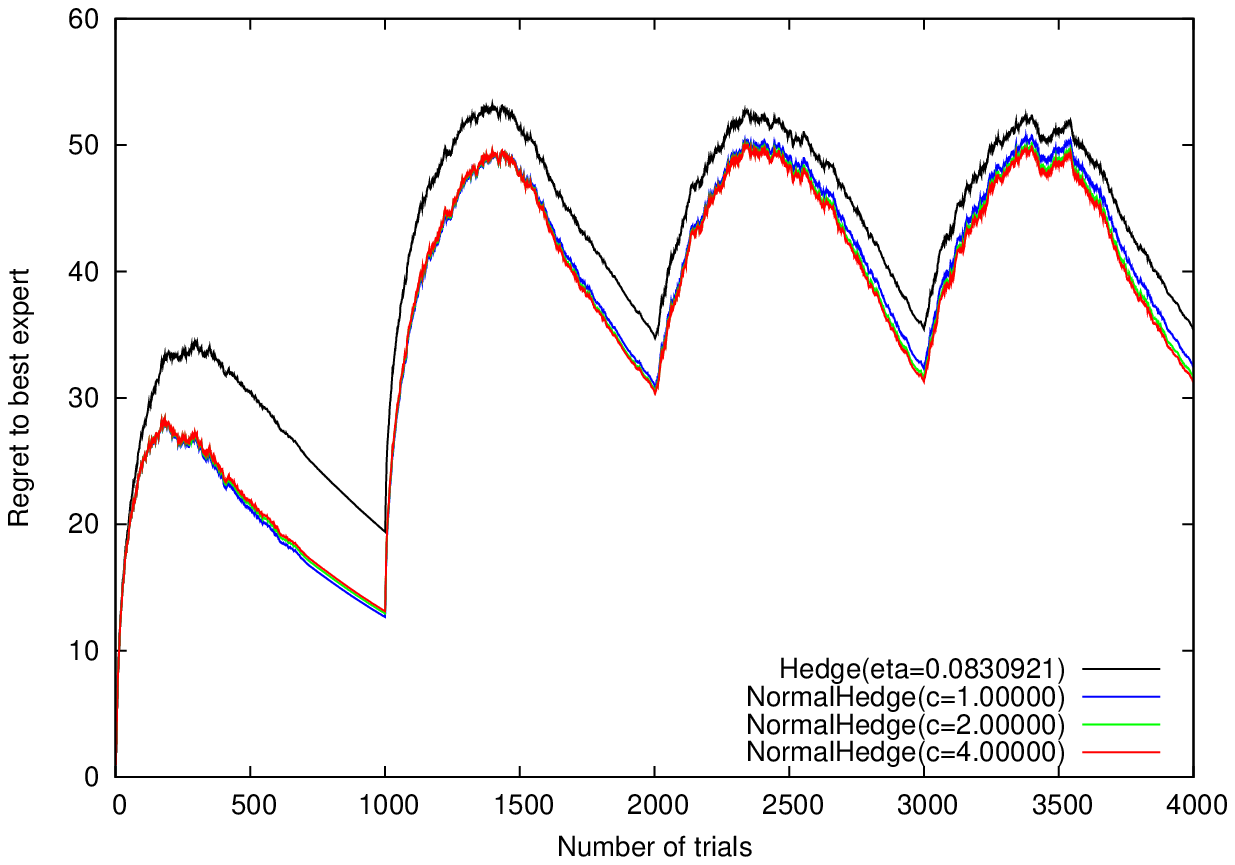} &
\includegraphics[width=0.43\textwidth]{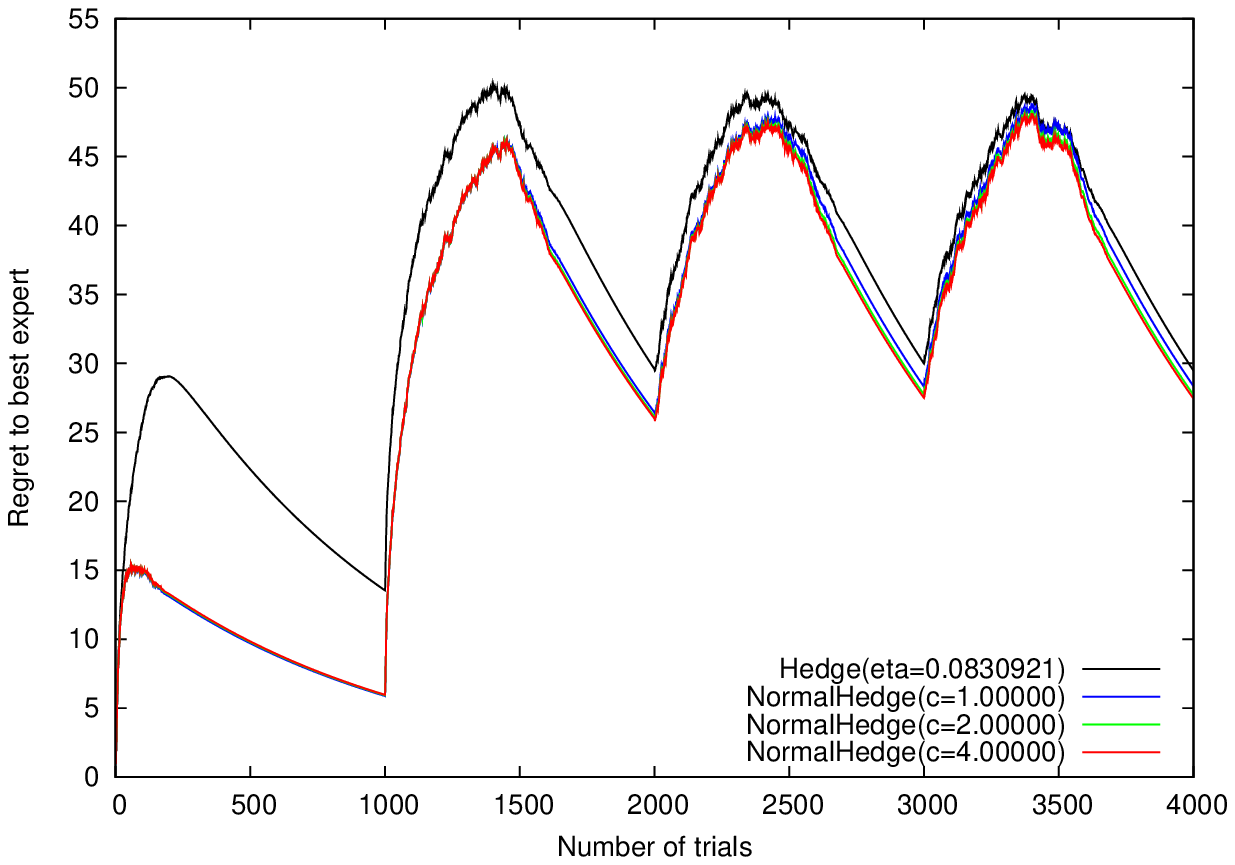} \\
$\gamma = 0.2, f = 0.1$ & $\gamma = 0.4, f = 0.1$ \\
\includegraphics[width=0.43\textwidth]{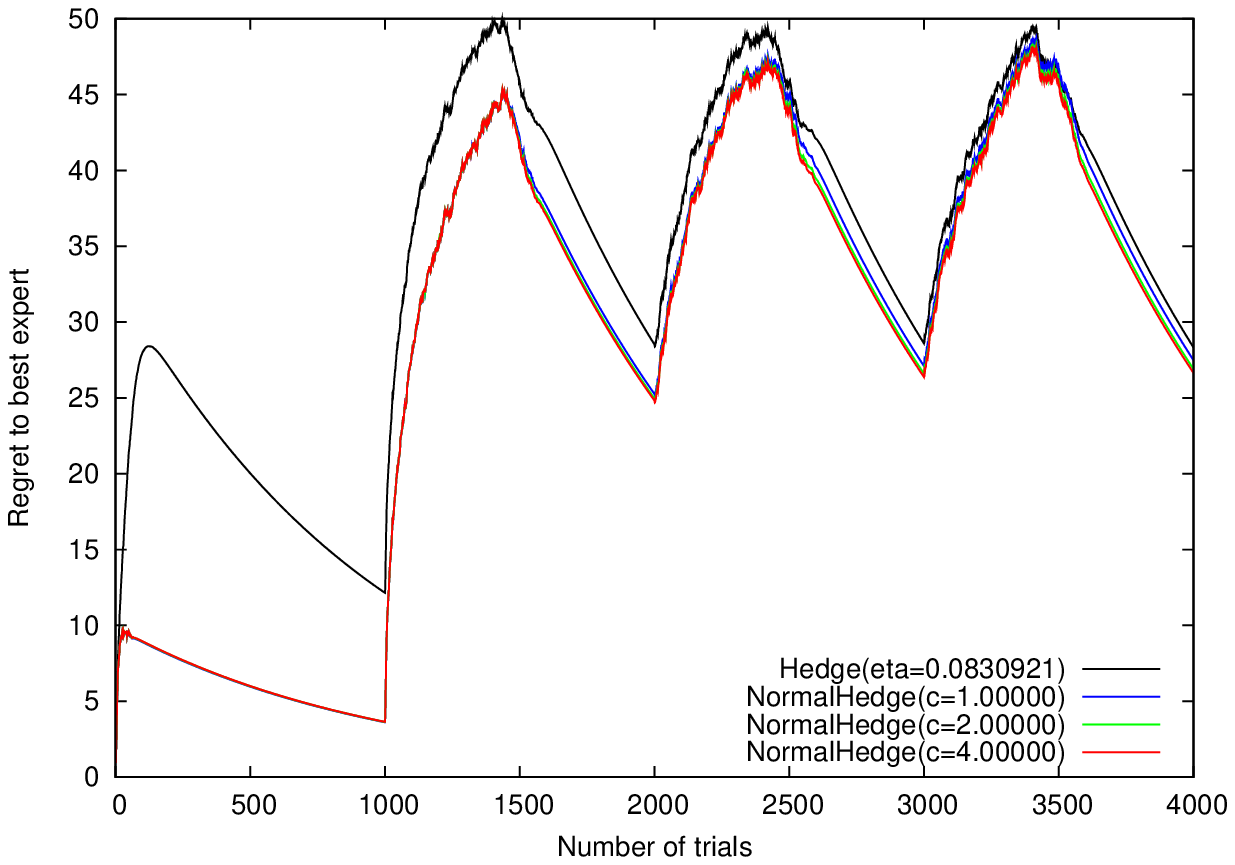} &
\includegraphics[width=0.43\textwidth]{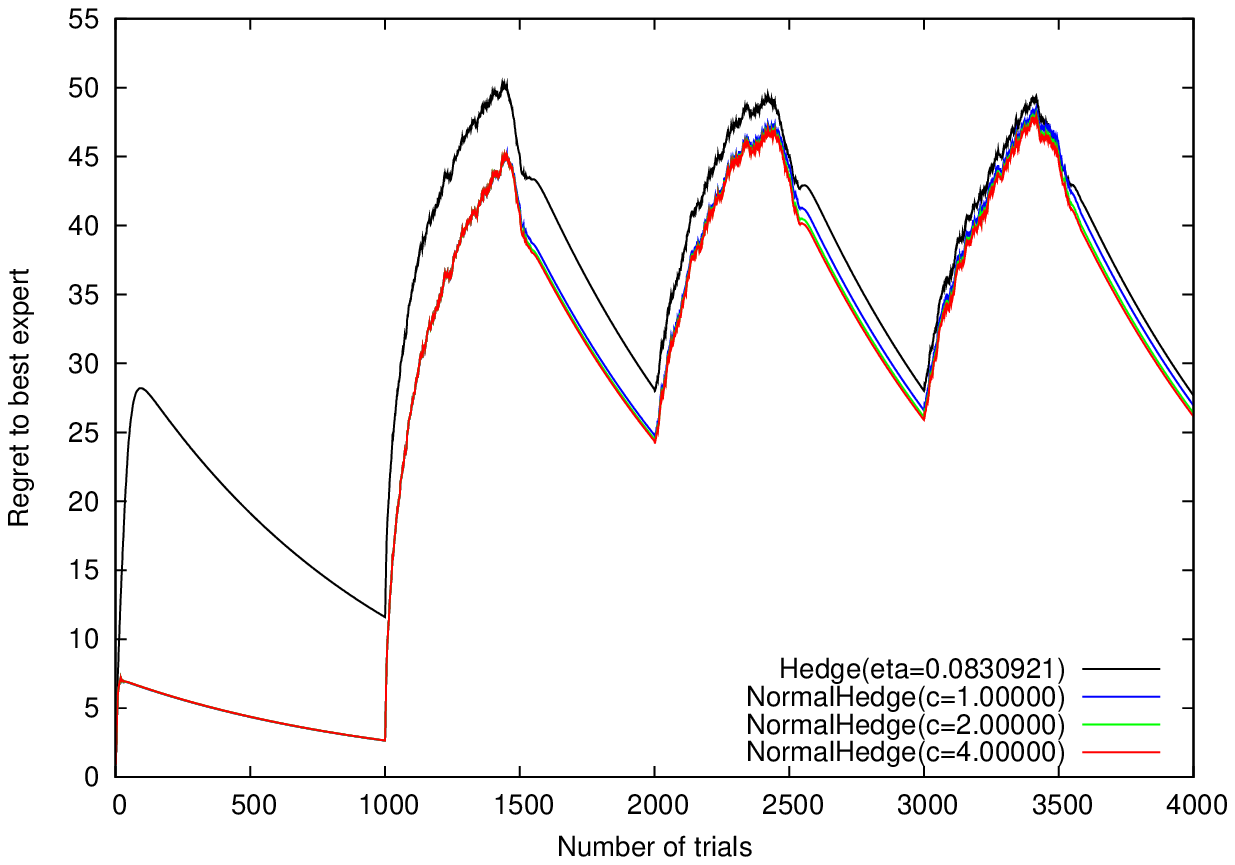} \\
$\gamma = 0.6, f = 0.1$ & $\gamma = 0.8, f = 0.1$ \\
\includegraphics[width=0.43\textwidth]{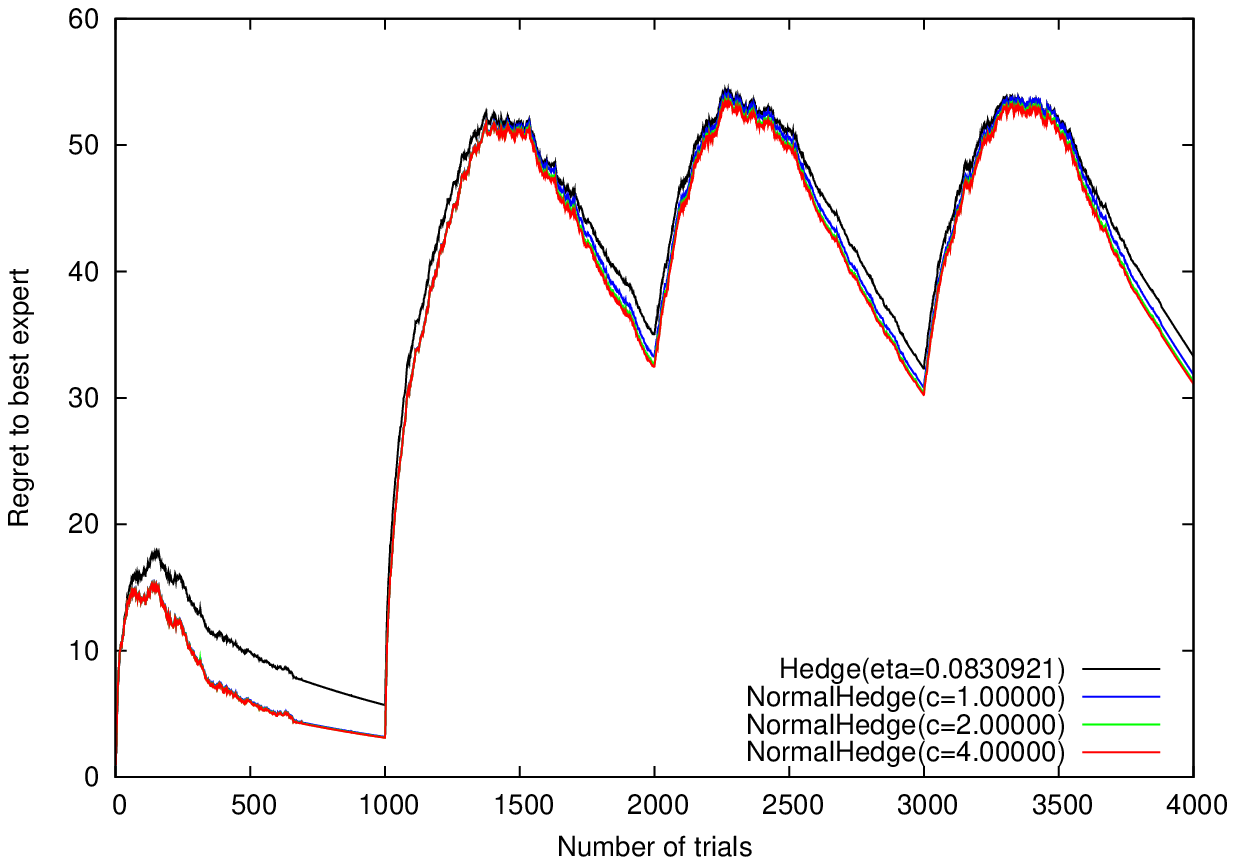} &
\includegraphics[width=0.43\textwidth]{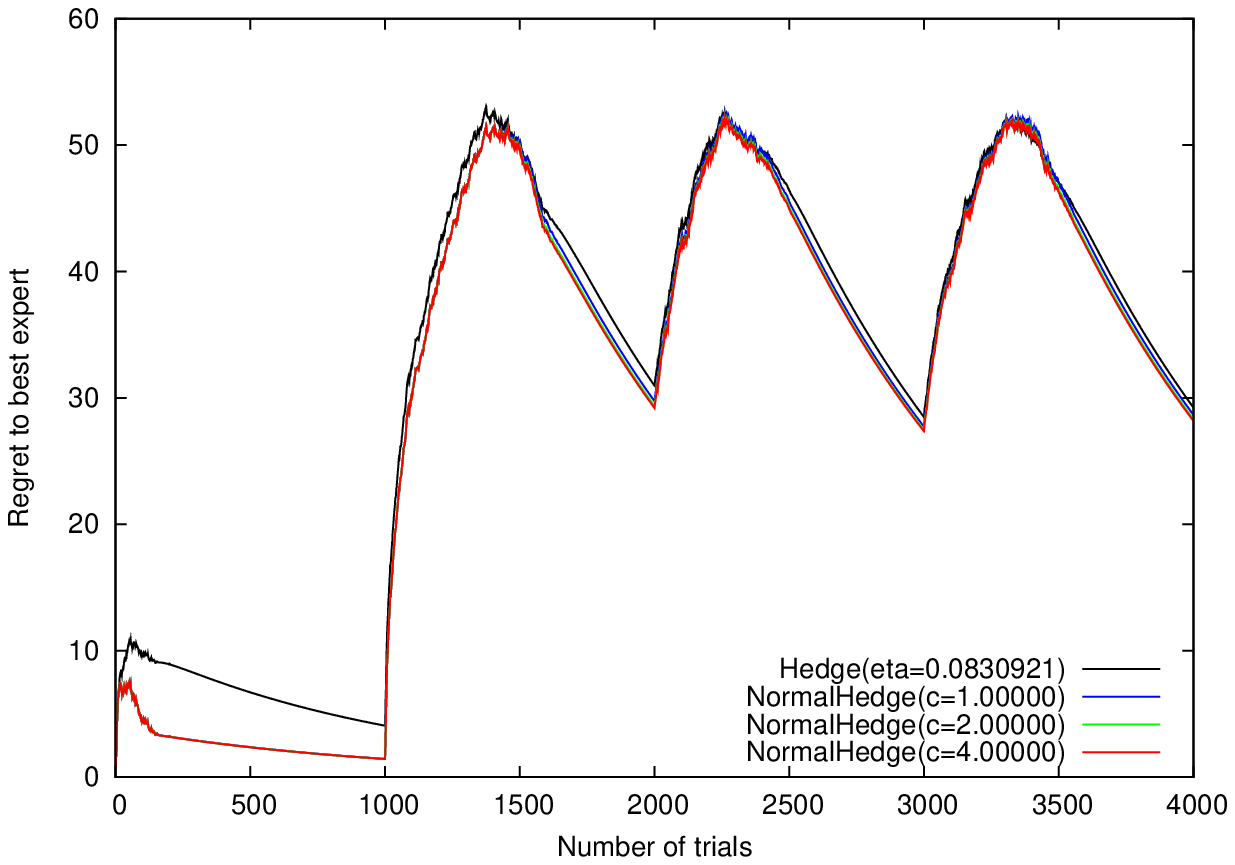} \\
$\gamma = 0.2, f = 0.5$ & $\gamma = 0.4, f = 0.5$ \\
\includegraphics[width=0.43\textwidth]{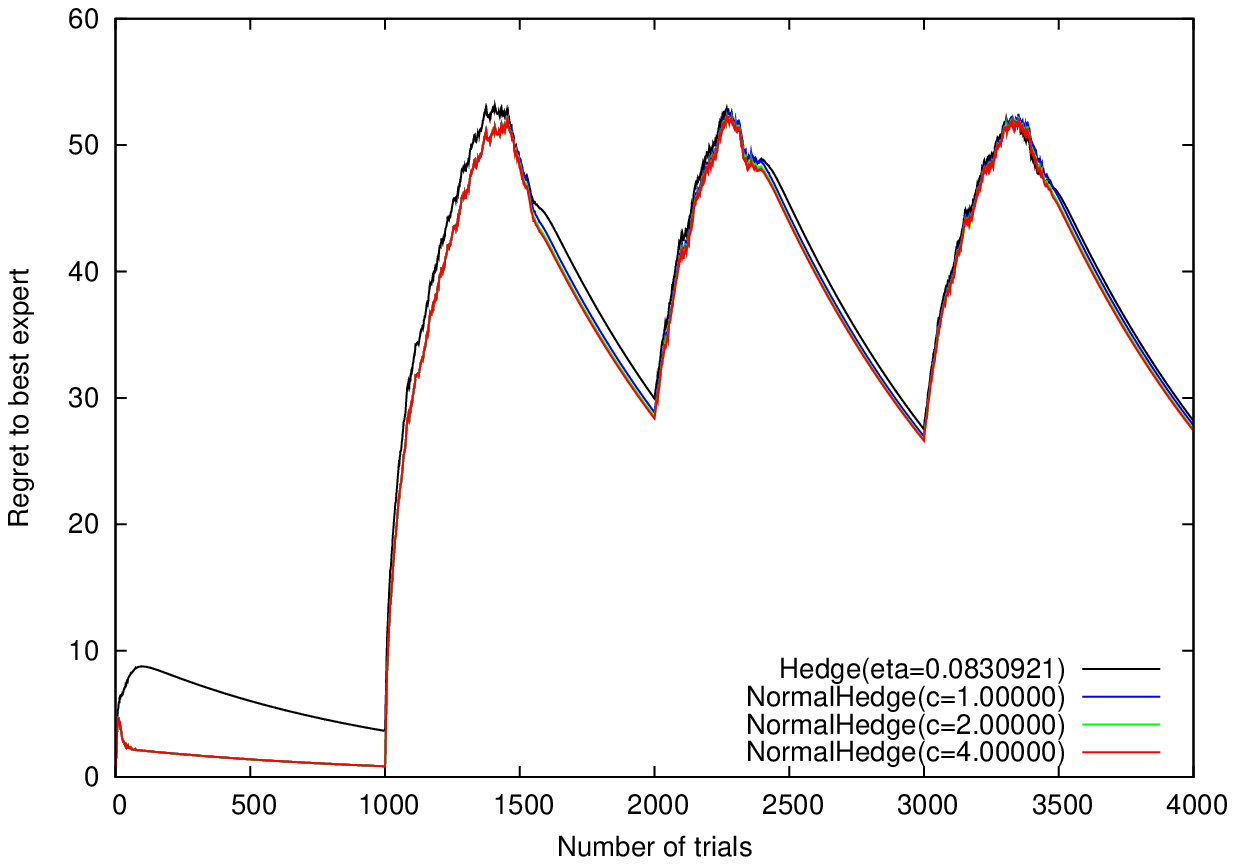} &
\includegraphics[width=0.43\textwidth]{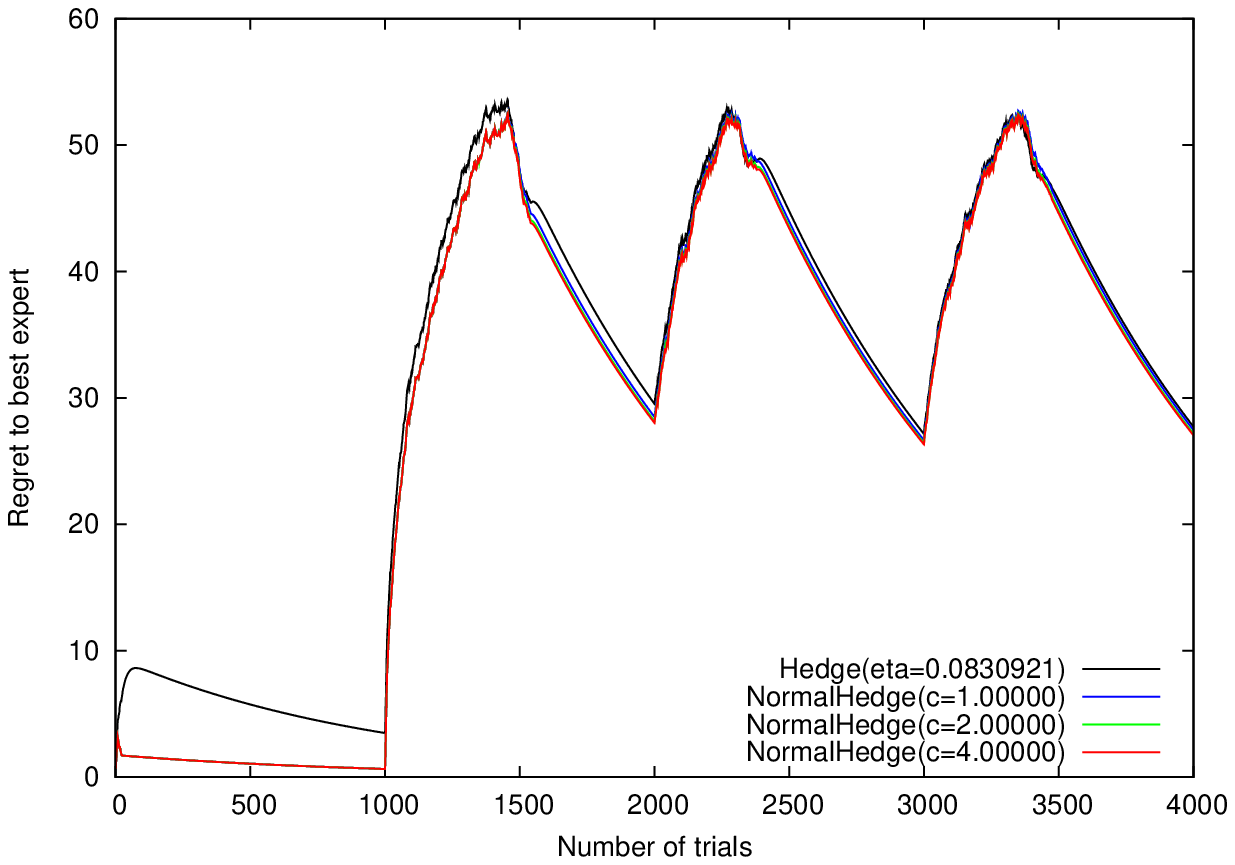} \\
$\gamma = 0.6, f = 0.5$ & $\gamma = 0.8, f = 0.5$
\end{tabular}
\end{center}
\caption{Regrets to the best expert in the first simulation;
$\gamma \in \{ 0.2, 0.4, 0.6, 0.8 \}$ and $f \in \{ 0.1, 0.5 \}$.}
\label{fig:sim1-2}
\end{figure}

Figures~\ref{fig:sim1-1} and \ref{fig:sim1-2} depict the discounted
cumulative regret to the best expert (averaged over $50$ runs). First, we
observe that NormalHedge fares better than Hedge when the advantage of the
good experts is large and the fraction of experts that are good is large.
In such cases, the advantage of NormalHedge is especially pronounced within
$1/\alpha$ iterations (before the set of good experts shifts). Second, we
observe that the performance of NormalHedge generally improves as the value
of $c$ is decreased. Indeed, the setting of $c = 1$ (for which we have no
theoretical guarantees) yields the best results for NormalHedge (and in
fact outperforms Hedge in every simulation). It would be very interesting
to establish guarantees for NormalHedge for $c \to 1$.

\subsubsection{The effect of tuning $\eta$ in Hedge}

Next, to bring out the issue with parameter tuning in Hedge, we conducted a
simulation in which we fix the fraction of experts that are good, but vary
the total number of experts:
\begin{itemize}
\item The number of experts is $N$, and the discount parameter is
$\alpha = 0.001$. (We varied $N \in \{ 10, 100, 1000 \}$.)
\item The fraction of experts that are good is fixed at $f = 0.1$. The
notion of good and bad experts is the same as in the first simulation. (We
varied $\gamma \in \{ 0.2, 0.8 \}$.)
\item The remaining details are the same as in the first simulation.
\end{itemize}
Again, we tuned the learning rate parameter for Hedge to $\eta =
\sqrt{(\alpha - \alpha^2/2) \log N}$, which now changes as we vary the
total number of experts, and we varied $c \in \{ 1, 2, 4 \}$ in
NormalHedge.

\begin{figure}
\begin{center}
\begin{tabular}{cc}
\includegraphics[width=0.43\textwidth]{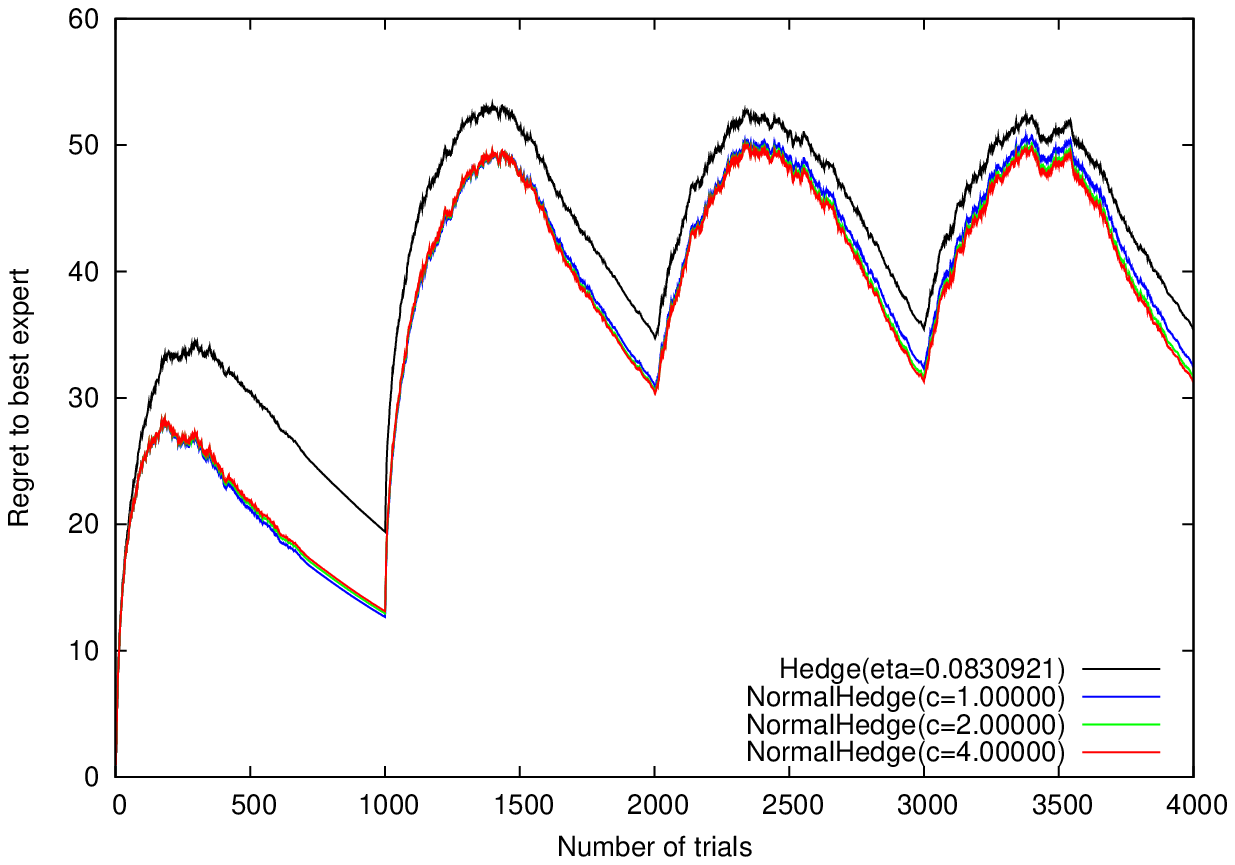} &
\includegraphics[width=0.43\textwidth]{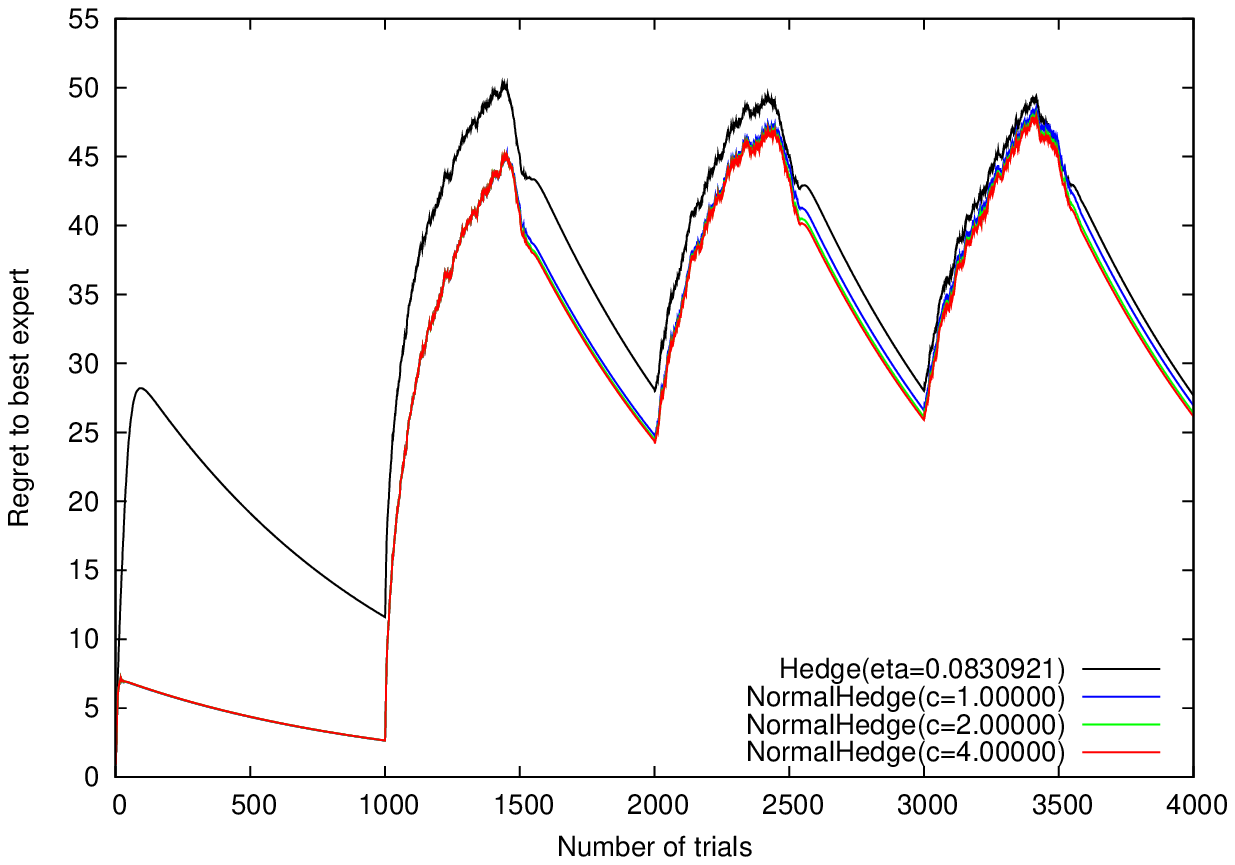} \\
$\gamma = 0.2, N = 1000$ & $\gamma = 0.8, N = 1000$ \\
\includegraphics[width=0.43\textwidth]{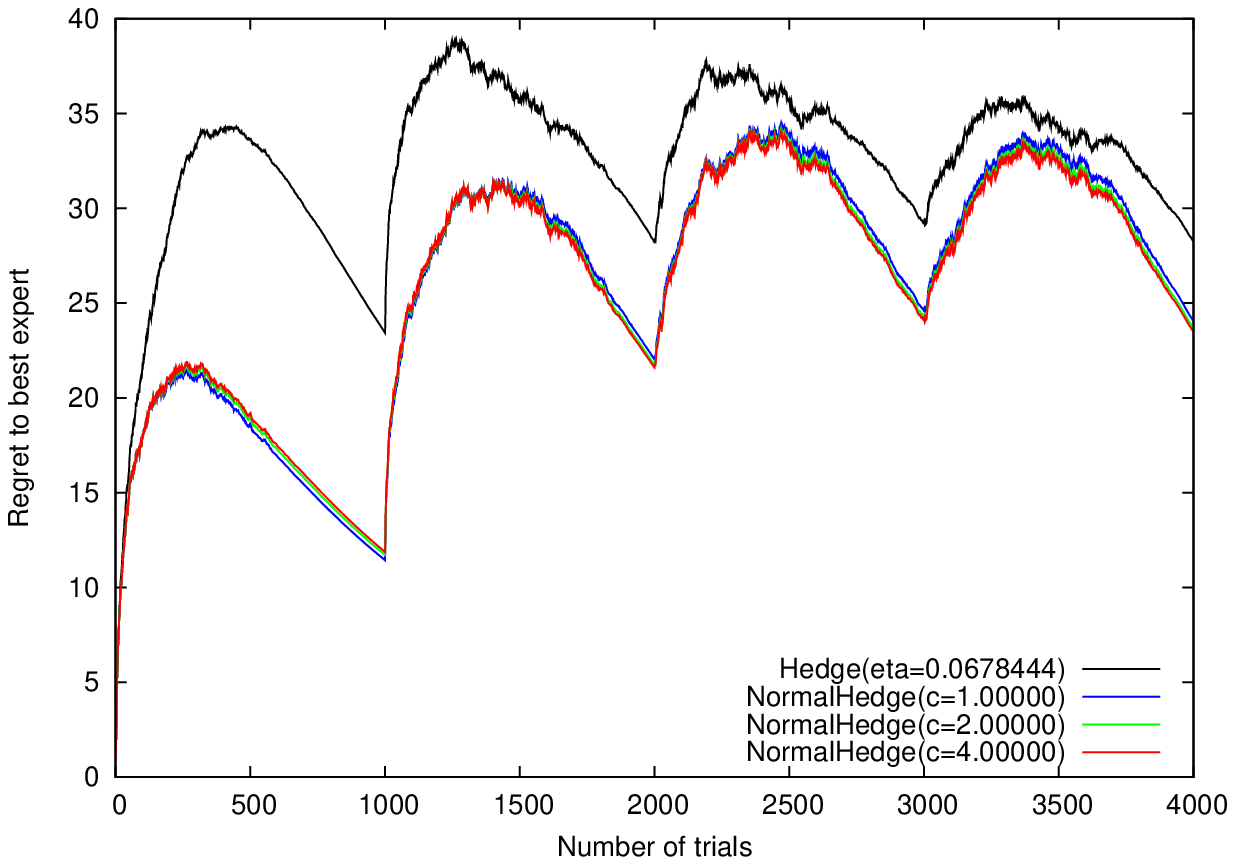} &
\includegraphics[width=0.43\textwidth]{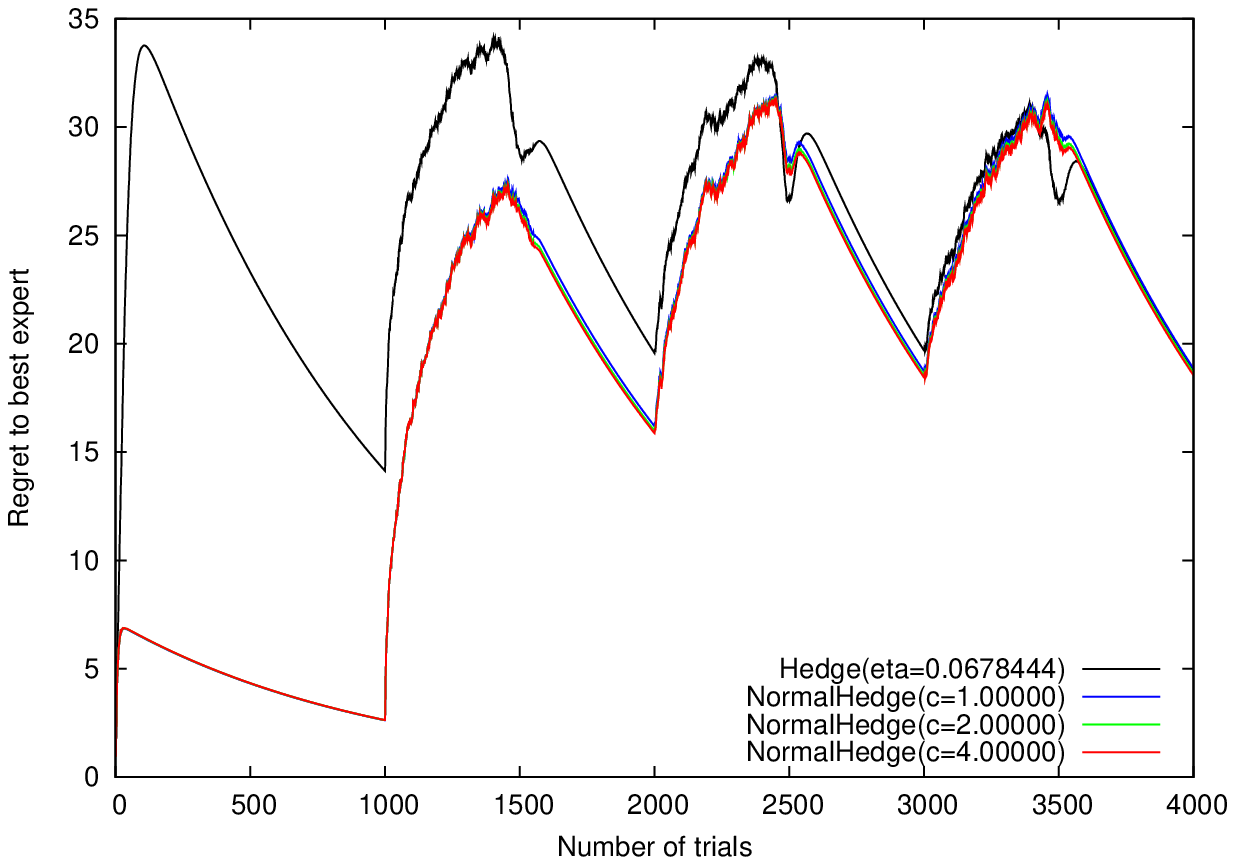} \\
$\gamma = 0.2, N = 100$ & $\gamma = 0.8, N = 100$ \\
\includegraphics[width=0.43\textwidth]{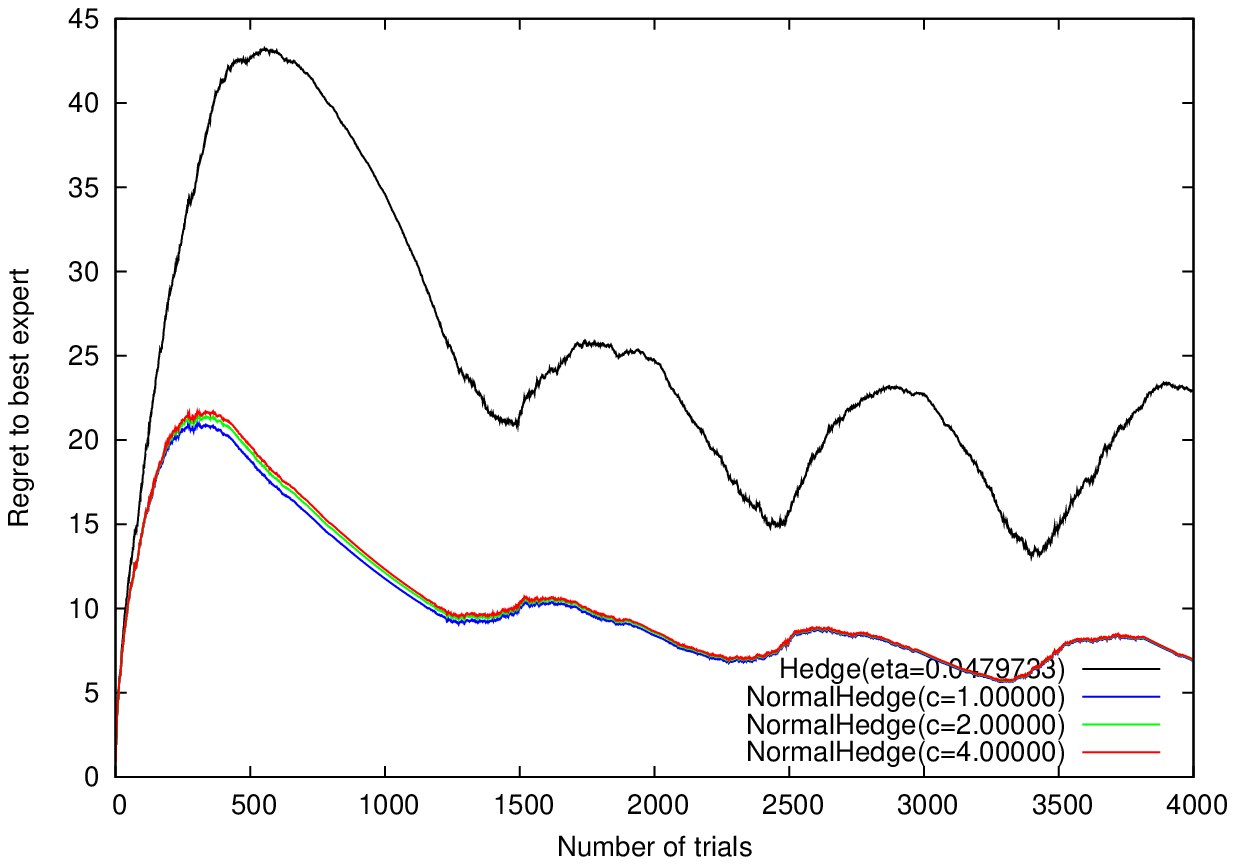} &
\includegraphics[width=0.43\textwidth]{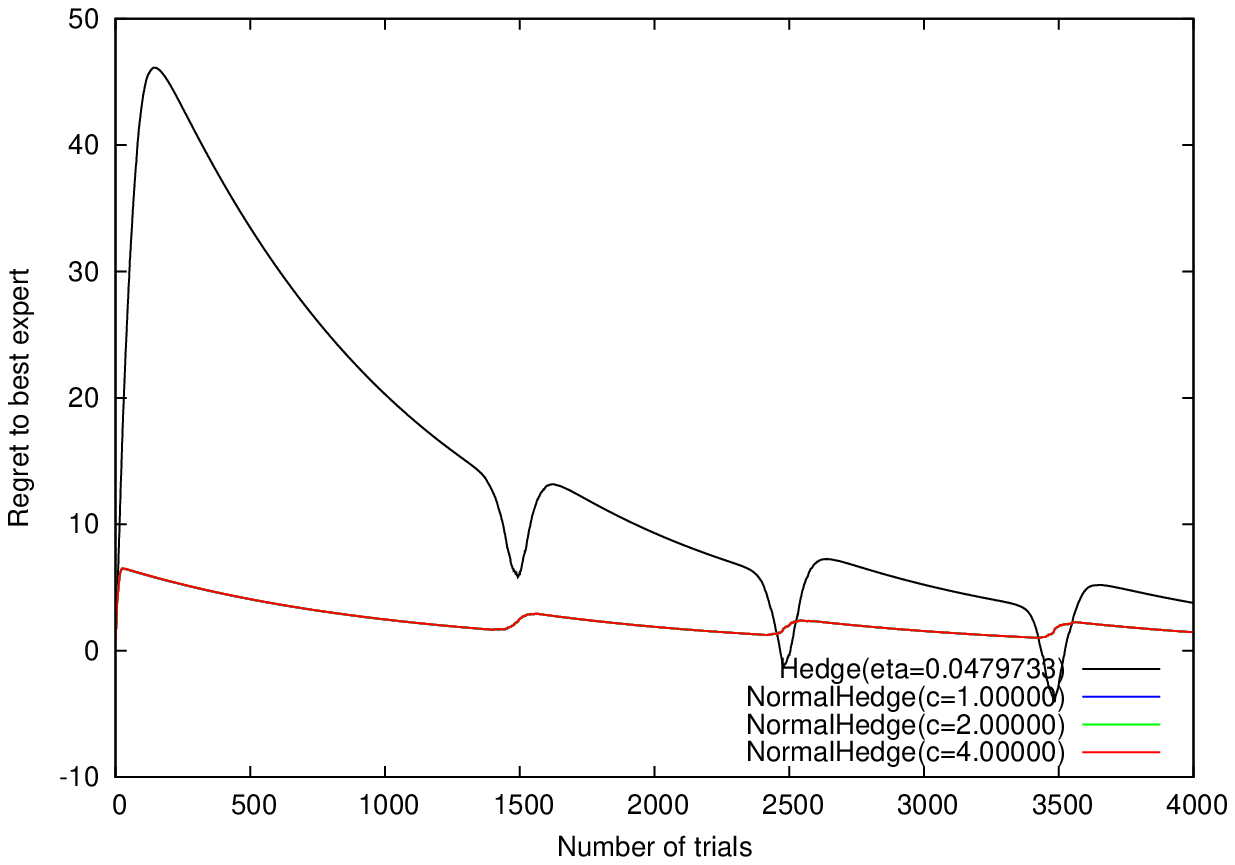} \\
$\gamma = 0.2, N = 10$ & $\gamma = 0.8, N = 10$
\end{tabular}
\end{center}
\caption{Regrets to the best expert in the second simulation;
$\gamma \in \{ 0.2, 0.8 \}$ and $N \in \{ 1000, 100, 10 \}$.}
\label{fig:sim2-1}
\end{figure}

The results (Figure~\ref{fig:sim2-1}) indicate that as $N$ decreases
(e.g.~$N = 100, 10$), the disparity between Hedge and NormalHedge
increases. We believe this is an issue with tuning the learning rate
$\eta$, which is conspicuously absent in NormalHedge, but we have not
precisely characterized the issue.

%
%
%
%
%
\section{Inferring latent random variables} \label{sec:latent} 

An important problem in statistical inference is to make predictions
or choose actions when the system under consideration has internal
states that cannot be observed directly. There are many manifestations
of this problem, including Graphical models, Hidden Markov Models
(HMMs), Partially Observable Markov Decision Processes (POMDPs) and
Kalman filters. The common method for dealing with hidden states is to
model them as {\em latent random variables}. The relation between the
latent random variables and the observable random variables is modeled
using a joint probability distribution. Two very important
sub-problems that arise in this approach are learning joint
distributions the involve latent random variables from examples that
contain only the state of the observable random variables and using
this type of joint distributions to infer the value of some
variables given the state of others. At this time there is no good
universal solution to either of these sub-problems.

We propose a different approach to the problem, where instead of
associating hidden states with hidden random variables, we associate
states with different experts. What we present here describes some
initial ideas. It is not an attempt to propose a solution to this
large and complex problem.

Suppose that we are to predict a binary sequence $x_1,x_2,\ldots$,
$x_t \in \{0,1\}$ and suppose that we believe that the sequence can be
predicted reasonably well using a Hidden Markov Model. Specifically,
suppose there is a hidden state $S$ which attains one of the values
$1,\ldots,k$ at each time step. Suppose that the state transition is
Markovian and stationary, i.e.
\[
P(S_t |S_{t-1},S_{t-2},\ldots) = P(S_t | S_{t-1}) = P(S_{t-1}|S_{t-2})
= \cdots
\]
Assume in addition that the hidden state does not change very often,
i.e. $P(S_{t+1} = S_t)$ is close to 1. Finally, assume that the
distribution of the observable variable $X_t$ depends only on the
hidden state at the same time $S_t$.

Consider the problem of predicting $X_{t+1}$ given $x_1,\ldots,x_t$
and the parameters of the HMM. Suppose that the prediction needs to
take the form of a distribution over $\Sigma$. So far this is exactly
the standard framework, but suppose we differ from the standard
framework by considering the $L_1$ loss $1-p_t(x_t)$, where $p_t(x_t)$
is the predicted probability assigned to the letter that actually
occured at time $t$. This is instead of the standard log likelihood
loss $\log(1/p_t(x_t)$. While the log loss is easier to analyze, the
$L_1$ loss is often a more useful measure because the cumulative $L_1$
loss corresponds to the expected number of mistakes. While this loss
does not fit well in the maximal likelihood or Bayesian methodologies,
it fits NormalHedge very well, because the loss per-iteration is bounded.

Here is our proposal for solving the prediction problem using
NormalHedge. We associate a set of experts with each hidden state. The
experts are confidence rated, i.e. each one of the experts outputs a
confidence level $0 \leq c \leq 1$ at each time step, the confidence
level is used in the confidence rated variant of NormalHedge described
in the previous section. If expert $i$ corresponds to a hidden state
$j$ then $c_i$ should be large when $S=j$ and low when $S \neq
j$. Suppose that the parameters of the HMM are known, then we can
associate a single expert with each hidden state and compute the
prediction and the confidence value of that expert using Bayes
formula.

Now suppose that we don't know the parameter vector of the HMM but
that we know that the vector is one of $N$ possibilities. In this case
we associate $N$ experts with each hidden state and compute the
predictions and confidence value of each expert using Bayes Formula
for the corresponding parameter vector, the confidence value for each
state is the a-posteriori probability for that state.

In this case the NormalHedge algorithm will quickly converge and give
most of the weight to the experts that correspond to the correct
parameter vector. Moreover, if none of the parameter vectors is
a correct description of the sequence distribution, it will converge
on the vector which causes the least regret, i.e. makes the smallest
number of mistakes.

Contrast this with the Bayesian approach. If the true distribution
generating the data is not included in the set of models over which we
take the posterior average, and if the loss function in which we are
interested is not log-likelihood but rather number of mistakes. Then
the cumulative loss of the Bayesian average can be much larger than
that of the best model in the set.

\section{Open problems}

The most interesting open problem is to close the gap between the
upper bound and lower bound on the parameter $c$. We have a lower
bound of $c>1$ and an upper bound of $c=4$. If we consider the case
$\alpha \to 0$ we can reduce $c$ to $2$. However, the gap between
$c=1$ and $c=2$ remains.

One promising direction of expansion is to consider the game in the
continuous time limit directly. This leads us naturally into
stochastic processes in continuous time such as Wiener
processes. Understanding the performance of NormalHedge in this
context might yield new methods for stochastic estimation and
stochastic control.

\bibliography{bib} \bibliographystyle{alpha}

\appendix

\section{Proof of main theorem}

Recall, the cumulative discounted regret of action $i$ at time $t=j\alpha$,
$j \in \N$ is defined recursively by
\[ R_i(0) = 0, \quad R_i(t+\alpha) = (1-\alpha) R_i(t) + g_i(t) - g_A(t),
\]
where $g_i(t) \in [-\sqrt{\alpha}, +\sqrt{\alpha}]$ is the (scaled) gain of
action $i$ at time $t$, and $g_A(t) \in [-\sqrt{\alpha}, +\sqrt{\alpha}]$
is the (scaled) gain of the hedger at time $t$. We define $r_i(t) = (g_i(t)
- g_A(t)) / \sqrt{\alpha} \in [-2,+2]$ as the (unscaled) instantaenous
regret to action $i$ at time $t$. The central quantity of interest is the
\emph{average potential}
\[ \Psi(t) = \frac1N \sum_{i=1}^N \Phi(R_i(t)). \]
Recall, we use the definition of the potential function $\Phi$
in Equation~\eqref{eqn:potential} with $c = 4$.

\begin{claim}
There exists a positive constant $C \leq 2.32$ such that if $\alpha <
1/(800\ln CN)$, then the average potential is always bounded from above by
$C$; that is, $\Psi(j\alpha) < C$ for any $j \in \N$.
\end{claim}
\begin{proof}
Fix $j \in \N$ and let $t = j\alpha$.

We will analyze the average $\Psi(t+\alpha) - \Psi(t)$ by considering
the averages over two separate groups:
\[ I_1 = \{ i : R_i(t) \leq 0 \} \quad \text{and} \quad I_2 = \{ i : R_i(t)
> 0 \}. \]

Let $\Psi_k(t) = (1/|I_k|) \sum_{i \in I_k} \Phi(R_i(t))$ be the average
potential for $I_k$, $k = 1, 2$ (assume without loss of generality that
neither $I_k$ is empty). We'll show the following facts:
\begin{enumerate}
\item[(A):] $\Psi_1(t) = 1$ and $\Psi_1(t+\alpha) < 1 + (3/5)\alpha$;
\item[(B):] If $\Psi(t) < 2.32$, then $\Psi_2(t+\alpha) - \Psi_2(t) < (2/3)\alpha$;
\item[(C):] If $2.31 < \Psi(t) < 2.32$, then $\Psi(t+\alpha) < \Psi(t)$.
\end{enumerate}
These facts imply that the increase in average potential from $\Psi(t)$ to
$\Psi(t+\alpha)$ is always less than $(2/3)\alpha < 1/1200$, and that if
the average potential $\Psi(t)$ is strictly between $2.31$ and $2.32$,
then $\Psi(t+\alpha)$ is strictly less than $\Psi(t)$. The claim then
follows by induction because $\Psi(0) = 1$.

We now prove the facts (A), (B), and (C).

(A): For $i \in I_1$, $\Phi(R_i(t)) = 1$ and $R_i(t+\alpha) \leq (1-\alpha)
R_i(t) + |r_i(t)|\sqrt{\alpha} \leq 2\sqrt{\alpha}$. Since $\Phi(x)$ is
non-decreasing in $x$, we have $\Phi(R_i(t+\alpha)) \leq
\Phi(2\sqrt{\alpha}) = e^{\alpha/2} < 1+\alpha/2+\alpha^2e^{\alpha/2}/2 <
1+(3/5)\alpha$ (the last inequality follows from the upper bound on
$\alpha$).

(B): We address terms in $I_2$ by expanding $\Phi(R_i(t+\alpha))$ around
the point $R_i(t) \neq 0$ via Taylor's theorem:
\[ \Phi(R_i(t+\alpha)) = \Phi(R_i(t)) + d_i(t) \Phi'(R_i(t)) + \frac12
d_i(t)^2 \Phi''(\rho_i) \]
where $d_i(t) = r_i(t) \sqrt{\alpha} - \alpha R_i(t)$ and $\rho_i \in \R$
lies between $R_i(t)$ and $R_i(t+\alpha)$. Because the hedger's weights are
chosen so that $p_i(t) \propto \Phi'(R_i(t))$, we have that
\[ \sum_{i=1}^N g_i(t) \Phi'(R_i(t)) - g_A(t) \sum_{i=1}^N \Phi'(R_i(t)) =
0 \]
and thus
\[ \Phi(R_i(t+\alpha)) - \Phi(R_i(t)) = -\alpha R_i(t) \Phi'(R_i(t)) +
\frac12 d_i(t)^2 \Phi''(\rho_i). \]
We need a few bounds before proceeding. First, if $\Psi(t) < 2.32$, then
$\Phi(R_i(t)) < 2.32N$ for all $i$, which implies $R_i(t) <
\sqrt{8\ln(2.32N)}$ for all $i$. By the condition on $\alpha$, we also have
$\sqrt{\alpha}R_i(t) < 1/10$. Next, we use a bound on $\rho_i$ since it is
evaluated in the non-decreasing function $\Phi''(x)$:
\[ \rho_i^2 \ \leq \ \max\{R_i(t), R_i(t+\alpha)\}^2 \ \leq
\ (R_i(t) + |r_i(t)|\sqrt{\alpha})^2
\ = \ R_i(t)^2 + 2\sqrt{\alpha}R_i(t)|r_i(t)| + \alpha r_i(t)^2
\ \leq \ R_i(t)^2 + \frac12. \]
Finally, we bound $d_i(t)^2$ as follows:
\[ d_i(t)^2 \ \leq \ (|r_i(t)|\sqrt{\alpha} + \alpha R_i(t))^2 \ \leq
\ (2+1/10)^2\alpha \ \leq \ \frac92 \alpha. \]
Altogether, we have
\begin{align*}
\Phi(R_i(t+\alpha)) - \Phi(R_i(t))
& = - \alpha R_i(t) \Phi'(R_i(t)) + \frac12 d_i(t)^2 \Phi''(\rho_i) \\
& = - \alpha R_i(t) \frac{R_i(t)}{4} e^{R_i(t)^2/8} + \frac12 d_i(t)^2
\left( \frac14 + \frac{\rho_i^2}{16} \right) e^{\rho_i^2/8} \\
& \leq - \alpha \frac{R_i(t)^2}{4} e^{R_i(t)^2/8} + \frac{9\alpha}{4}
\left( \frac9{32} + \frac{R_i(t)^2}{16} \right) e^{R_i(t)^2/8} e^{1/16} \\
& \leq \alpha \left(\frac23 - \frac1{10} R_i(t)^2\right) e^{R_i(t)^2/8}.
\end{align*}
The final bound is decreasing as a function of $R_i(t) \geq 0$. This
implies $\Phi(R_i(t+\alpha)) - \Phi(R_i(t)) \leq (2/3)\alpha$, so
$\Psi_2(t+\alpha) - \Psi_2(t) < (2/3)\alpha$.

(C): First, consider the problem of maximizing
\[ f(x_1, \ldots, x_n) = \sum_{i=1}^n \left( \frac23 - \frac{x_i^2}{10}
\right) e^{x_i^2/8} \]
subject to the constraint $(1/n) \sum_{i=1}^n e^{x_i^2/8} \geq B$ for some
$B \geq 1$. Simple variational arguments imply that the maximum is attained
when $x_i = \sqrt{8\ln B}$ for all $i$. Therefore, following the argument
for (B), we have that if $\Psi_2(t) \geq B$ for some $B \geq 1$, then
\[ \Psi_2(t+\alpha) - \Psi_2(t) \leq \alpha \cdot B \cdot \left( \frac23 -
\frac45 \ln B \right). \]

Let $p_1 = |I_1|/N$ and $p_2 = 1 - p_1$. Suppose
$\Psi(t) > 2.31$. Because $\Psi_1(t) = 1$, we have
\[ \Psi_2(t)
= \frac1{p_2} \left( \Psi(t) - p_1 \right) \geq \frac1{p_2} (2.31 - p_1)
\doteq B. \]

Now we analyze the overall change in average potential. By (A), the
increase in average potential over $i \in I_1$ is less than
$(3/5)\alpha$. Then
\begin{align*}
\frac{\Psi(t+\alpha) - \Psi(t)}{\alpha}
& < p_1 \cdot \frac35 + p_2 \cdot B \cdot \left( \frac23 - \frac45 \ln B
\right) \\
& = p_1 \cdot \frac35 + (2.31 - p_1) \cdot \left( \frac23 - \frac45 \ln
\frac1{1-p_1} - \frac45 \ln (2.31 - p_1) \right).
\end{align*}

The final RHS is decreasing as a function of $p_1 \geq 0$, so it is
maximized when $p_1 = 0$. Making this substitution, the RHS is negative,
and thus $\Psi(t+\alpha) < \Psi(t)$.
\end{proof}

\end{document}